\documentstyle[epsf]{elsart}

\begin{document}

\begin{frontmatter}

\title{Stability of Oscillating Hexagons in Rotating Convection}

\author{Blas Echebarria\thanksref{Co}}
\author{\and Hermann Riecke}

\address{
Department of Engineering Sciences and Applied Mathematics,
Northwestern University, 2145 Sheridan Rd, Evanston, IL, 60208, USA
}

\thanks[Co]{Corresponding author. Tel.: (847) 467-3451, Fax.: (847) 491-2178, email: blas@korf.esam.nwu.edu.}

\begin{abstract}

Breaking the chiral symmetry, rotation induces a secondary Hopf bifurcation in
weakly nonlinear hexagon patterns which gives rise to oscillating hexagons. We study the stability of the oscillating hexagons using three coupled 
Ginzburg-Landau equations. Close to the bifurcation point we derive
reduced equations for the amplitude of the oscillation, coupled to the phase
of the underlying hexagons. Within these equation we identify two types of
long-wave instabilities and study the ensuing dynamics using numerical
simulations of the three coupled Ginzburg-Landau equations.

\end{abstract}



\end{frontmatter}

\section{Introduction}

Convection has played a key role in the elucidation of the 
spatio-temporal dynamics arising in non-equilibrium pattern forming 
systems.  The interplay of well-controlled experiments with analytical 
and numerical theoretical work has contributed to a better 
understanding of various mechanisms that can lead to complex behavior.  
From a theoretical point of view the effect of rotation on roll 
convection has been particularly interesting because it can lead to 
spatio-temporal chaos immediately above threshold where the small 
amplitude of the pattern allows a simplified treatment.  Early work of 
K\"{u}ppers and Lortz \cite{KuLo69,Ku70} showed that for sufficiently 
large rotation rate the roll pattern becomes unstable to another set 
of rolls rotated with respect to the initial one.  Due to isotropy the 
new set of rolls is also unstable and persistent dynamics are 
expected.  Later Busse and Heikes \cite{BuHe80} confirmed experimentally the 
existence of this instability and the persistent dynamics arising from 
it.  They proposed an idealized model of three coupled 
amplitude equations 
in which the instability leads to a heteroclinic cycle connecting 
three sets of rolls rotated by 120$^o$ with respect to each other.  
Recently the K\"uppers-Lortz instability and the ensuing dynamics have 
been subject to intensive research, both experimentally 
\cite{ZhEc91,ZhEc92,NiEc93,HuEc95,HuPe98} and theoretically 
\cite{XiGu94,TuCr92,FaFr92,NeFr93,ClKn93,CrMe94,PoPa97}.  It is 
found that in sufficiently large systems the switching between rolls 
of different orientation looses coherence and the pattern breaks up 
into patches in which the rolls change orientation at different times.  The shape and 
size of the patches changes persistently due to the motion of the 
fronts separating them. 

In this paper we are interested in the effect of rotation on hexagonal 
rather than roll (stripe) patterns as they arise in systems with 
broken up-down symmetry (e.g.  non-Boussinesq or 
surface-tension driven convection with rotation).  
The dynamics of strictly periodic hexagon patterns with broken chiral 
symmetry have been investigated 
in detail by Swift \cite{Sw84} and Soward \cite{So85}. They found 
that the heteroclinic orbit of the Busse-Heikes model is replaced by a 
periodic orbit arising from a secondary Hopf bifurcation off the 
hexagons.  Their results have been confirmed in numerical simulations 
of a Swift-Hohenberg-type model \cite{MiPe92} in which an alternation 
among the three modes that compose the hexagonal pattern is observed. 
In the following we 
will call this state `oscillating hexagons'. The oscillations can be 
homogeneous in space or can take on the form of traveling waves. Starting from 
coupled 
Ginzburg-Landau equations, we have previously derived evolution 
equations for the oscillating hexagons that are valid close to the 
Hopf bifurcation. Within this framework the 
oscillating hexagons were found to support a state of spatio-temporal 
chaos that is characterized by defects, 
in which the oscillation amplitude vanishes \cite{EcRiunpub}. 
At the band center the 
oscillating hexagons and their chaotic state are 
described by the single complex Ginzburg-Landau 
equation (CGLE). In general, however, the oscillation amplitude is coupled to 
the phases of the underlying pattern as it is to be expected 
for a secondary bifurcation. Here we study how this coupling affects the stability of the oscillations. In particular, it will modify the stability properties
of the waves emitted by the spirals in the defect chaotic state. We find that the additional coupling  
 leads to new long-wave instabilities if the rotation is strong enough 
or if the wavenumber of the hexagons is far away from the band-center.\footnote{For stripe patterns (e.g. convection rolls) the generic equation 
for a secondary Hopf bifurcation is well known \cite{CoIo90}, and has been 
studied both theoretically and experimentally \cite{DaLe92,Sa93,JaKo92}. 
There, the  coupling to the phase of the pattern can delay the occurrence of 
long-wave instabilities of the oscillatory mode.}

The paper is organized as follows. In the second section we introduce 
the appropriate coupled Ginzburg-Landau equations that describe the 
hexagon patterns and derive the 
amplitude-phase equations close to the secondary Hopf bifurcation. 
The stability analysis of these equations is addressed in section III. 
In section
IV we study numerically the behavior resulting from these instabilities. 
Conclusions are given in section V. 
Details of the calculations are given in two appendices. 

\section{Amplitude-Phase equations for Oscillating Hexagons}

We consider small-amplitude hexagon patterns in systems with broken 
chiral symmetry. In order to analyze the possibility of modulational instabilities we include spatial derivatives. Due to the strong 
coupling between modes of different orientation, we take 
the gradients in both directions to be of the same order and retain only 
linear gradient terms (a study of the influence of nonlinear gradient terms
in the stability of oscillating hexagons is addressed in appendix B).  After rescaling the amplitude, time, and space we arrive at the equations,
\begin{eqnarray}
\partial _{t}A_{1} & = & \mu A_{1}+({\bf {n}}_{1}\cdot \nabla )^{2}A_{1}+
\overline{A}_{2}\overline{A}_{3} -  A_{1}|A_{1}|^{2} \label{eq.amp} 
 \\
&&-(\nu +\gamma)A_{1}|A_{2}|^{2}-(\nu -\gamma)A_{1}|A_{3}|^{2},\nonumber 
\end{eqnarray}
where the equations for the other two amplitudes are obtained by 
cyclic permutation of the indices and $\mu$ is a parameter 
related to the distance from threshold. The overbar represents complex conjugation. These equations can be 
obtained from the corresponding physical equations (e.g. 
Navier-Stokes) using a 
perturbative technique. The broken chiral 
symmetry  manifests itself by the cross-coupling coefficients not being
equal. Hence $\gamma$ is a measure of rotation.

For completeness it should be noted that rotation leads in 
convection not only to a chiral symmetry breaking but 
also to a (weak) breaking of the translation symmetry due to the 
centrifugal force. In the following we will 
consider it to be negligible.  We focus on not too small Prandtl numbers,
in which case the primary bifurcation is always steady \cite{Ch61,ClKn93}.

\begin{figure}
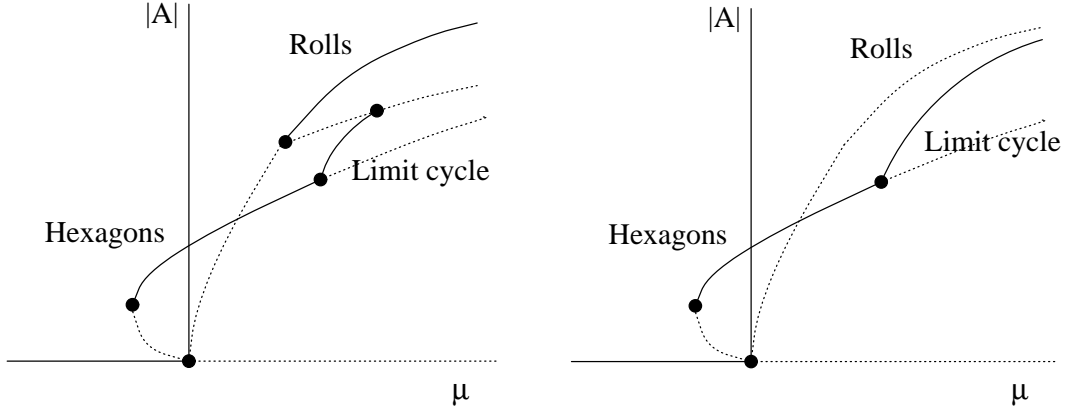

\centerline{\epsfxsize=6.5cm\epsfbox{figuras/bif1.eps}
\hspace{1cm}\epsfxsize=6.5cm\epsfbox{figuras/bif2.eps}}
\caption{Sketch of a bifurcation diagram for a fixed value of the wavenumber $q$. a) For $\gamma < \gamma_{{KL}}$ limit cycle of oscillating hexagons 
ends via heteroclinic cycle and rolls become stable. 
b) For $\gamma > \gamma_{KL}$ oscillating hexagons persist and rolls 
do not become stable.
\label{fig.bifos}}
\end{figure}

Equation (\ref{eq.amp}) admits hexagon solutions 
 \( A_{j}=Re^{iq\hat{\bf n}_{j}\cdot {\bf {x}}+i\phi_j} \)  with a slightly off-critical 
wavenumber ($\tilde{\bf q}_j=\tilde{\bf q}^c_j + {\bf q}_j$, $|{\bf q}_j | \ll | \tilde{\bf q}_c|\equiv \tilde{q}_c$), with
\begin{equation}
R=\frac{1\pm \sqrt{1+4(\mu -q^{2})(1+2\nu )}}{2(1+2\nu )}, \qquad \Phi\equiv \phi_1+\phi_2+\phi_3=0.\label{eq.hexa}
\end{equation}
The stability of this solution to perturbations with the same 
wave vectors has been studied by 
several authors \cite{Sw84,So85}. Typical results are sketched in the 
bifurcation diagrams shown in Fig.  \ref{fig.bifos}.  The hexagons appear 
through a saddle-node bifurcation at $\mu=\mu_{sn}$,
\begin{equation}
\mu_{sn} = -\frac{1}{4(1+2\nu)} + q^2,
\end{equation}
 and become unstable via a Hopf bifurcation at \( \mu =\mu_{H} \), with a frequency $\omega_H$,  
\begin{equation}
\mu_H = \frac{(2+\nu)}{(\nu-1)^2} + q^2,\;\; \omega_H=2\sqrt{3}\gamma/(\nu -1)^{2}\label{eq.trOH}.
\end{equation}
The Hopf bifurcation is supercritical and for $\mu > 
\mu_H$  stable oscillations in the three amplitudes of the 
hexagonal pattern arise with a phase shift of $2\pi/3$ between them 
\cite{Sw84,So85,MiPe92}, resulting in what we call 
oscillating hexagons.  As  \( \mu \) is increased further, eventually a 
point $\mu=\mu_{het}$ is reached at which the branch of oscillating hexagons ends on the branch corresponding to a mixed-mode solution in a global bifurcation involving a heteroclinic connection (see Fig. \ref{fig.bifos}a).  Above this point the only stable 
solution is the roll solution whose stability region is bounded below by
\begin{equation}
\mu _{R}=\frac{1}{(\nu +\gamma -1)(\nu -\gamma -1)}+q^{2}. \label{eq.trR}
\end{equation}
From Eqs. (\ref{eq.trOH}), (\ref{eq.trR}) it is easy to see that, when
$\gamma^2 > (\nu-1)(\nu+1)/(\nu+2)$,
the transition to oscillating hexagons occurs at a value of $\mu$ for which
the rolls are still unstable. There is then a 
parameter regime in which the oscillating hexagons are the only stable solution.  Furthermore, when $|\gamma| > \nu-1\equiv \gamma_{KL}$ rolls are never stable and the 
limit cycle persists for arbitrarily large values of $\mu$ (see Fig. \ref{fig.bifos}b). In the absence 
of the quadratic terms in Eq. (\ref{eq.amp}) this condition corresponds to the 
K\"uppers-Lortz instability of rolls. When the quadratic term in Eq. (\ref{eq.amp}) is small (i.e. small non-Boussinesq effects) it can be 
considered as a perturbation of the usual three mode model for rotating roll
convection. Far above the Hopf bifurcation ($\mu \gg \mu_H$) the periodic orbit
is expected to become asymmetrical and the resulting state similar to that 
encountered in the usual rotating Rayleigh-B\'enard convection.

The stability of steady hexagons with respect to side-band perturbations in the 
presence
of rotation has been studied previously \cite{EcRipre}. Due to the rotation 
steady and oscillatory, long-wave and short-wave instabilities have been found.
It turns out that in the 
presence of rotation the steady hexagons can be stable up to the Hopf 
bifurcation over quite a range of wavenumbers. Thus, the oscillating hexagons 
may be stable near the Hopf bifurcation.

The focus of the present paper are
side-band instabilities of the oscillating hexagons. To address this question 
analytically we focus on the vicinity
of the Hopf bifurcation where the oscillation amplitude is small and  
a weakly nonlinear analysis is possible. Since we are dealing with a secondary 
bifurcation the two
phases of the underlying hexagons have to be taken into account as well. In fact, from a linear stability analysis 
of Eq. (\ref{eq.amp}) it is easy to see that there are four marginal modes at
$\mu=\mu_H$. Two correspond to the Hopf bifurcation and can be described by a
complex amplitude ${\cal H}$. The other two correspond to a two-dimensional phase
vector $\vec{\phi}$, related to translations in the x- and y-directions \cite{LaMe93,Ho95}. It can be written as a combination of the phases of the 
three modes: $\vec{\phi}=(\phi_x, \phi_y)\equiv (-\phi_2-\phi_3,(\phi_2-\phi_3)/\sqrt{3})$, satisfying the 
locking condition $\phi_1 + \phi_2 + \phi_3 = \Phi$. The global phase $\Phi$ is
a fast variable that relaxes rapidly to its stationary values $\Phi=0,\pi$ (up or down hexagons). To study the nonlinear behavior of the oscillating hexagons, the amplitudes $A_i$ are expanded as:
\begin{equation}
A_n=(R_H+ e^{2\pi ni/3}\sqrt{\epsilon}{\cal H}e^{i\omega_H t} 
+ c.c. + {\cal O}(\epsilon) )e^{i{\bf q}_i \cdot {\bf x} +\sqrt{\epsilon}\phi_i},
\label{eq.expan}
\end{equation}
where $\epsilon$ is a small parameter related to the distance from the 
bifurcation line. The amplitude of the steady hexagons at the bifurcation point, $R_H=1/(\nu-1)$, is independent of $\mu$ and $q$. Eliminating the fast variables, at order $\epsilon^{3/2}$ (see appendix A) we
arrive at an equation for the amplitude of the oscillation ${\cal H}$, coupled to the
phase vector of the underlying hexagonal pattern,
\begin{eqnarray}
\partial_T{\cal H}&=&\varepsilon\delta_1{\cal H}+\xi\nabla^2{\cal H}-\delta_2{\cal H}
\nabla\cdot\vec{\phi}-\rho{\cal H}|{\cal H}|^2, \label{eq.cgle-ph.a}\\
\partial_T \vec{\phi}&=&D_{\bot}\nabla^2 \vec{\phi}+D_{\|}\nabla(\nabla\cdot
\vec{\phi})+D_{\times_1}({\bf e}_z\times\nabla^2\vec{\phi}) \label{eq.cgle-ph.b}
\\
&&+D_{\times_2}({\bf e}_z\times\nabla)(\nabla\cdot\vec{\phi})+\alpha\nabla|{\cal H}|^2 +\beta_1({\bf e}_z\times\nabla)|{\cal H}|^2 \nonumber \\
&&-i\beta_2({\cal H}\nabla\overline{{\cal H}} - \overline{{\cal H}}\nabla{\cal H})+i\eta[{\cal H}({\bf e}_z\times\nabla)
\overline{{\cal H}}-\overline{{\cal H}}({\bf e}_z\times\nabla){\cal H}], \nonumber
\end{eqnarray}
with $\varepsilon=\mu-\mu_H$ and
\begin{eqnarray}
&&v=3R_H(1+2R_H), \label{coeffs.1}\\
&&\delta_1=\frac{2R_H}{v}-\frac{2i\omega_H}{v},\;\;\delta_2=q\delta_1,\\
&&\xi=\frac{1}{2} - \frac{3q^2 R_H}{9R_H^2 + \omega_H^2} - \frac{iq^2}{\omega_H}\frac{9R_H^2+2\omega_H^2}{9R_H^2 + \omega_H^2},\\ 
&&\rho=\frac{8(3R_H+1)}{v}-\frac{4i\omega_H (1+4R_H)}{R_Hv} -\frac{32i}{3\omega_H},\\
&&D_{\bot}=\frac{1}{4},\;\;D_{\|}=\frac{1}{2} -\frac{2q^2}{v},\;\;D_{\times_1}=\frac{q^2}{\omega_H},\;\;D_{\times_2}=0,\\
&&\alpha=-\frac{2\omega_H^2 q}{9R_H^2 + \omega_H^2} - \frac{2q(1+6R_H)}{R_Hv},\\
&&\beta_1=\frac{6\omega_H q}{R_H(9R_H^2+\omega_H^2)},\;\;\beta_2=\beta_1,\;\;\eta=\frac{18q}{9R_H^2+\omega_H^2}
\label{coeffs.2}.
\end{eqnarray}
 
It is worth pointing out that the phase-amplitude equations (\ref{eq.cgle-ph.a},\ref{eq.cgle-ph.b}) can be deduced by means of
symmetry arguments alone and are, therefore, generic to this order in 
$\epsilon$. In fact, they could be derived directly from the fluid equations without the use of the Ginzburg-Landau equations (\ref{eq.amp}). Thus, keeping higher order terms in  (\ref{eq.amp})  would change the values of the coefficients, but not their form. 

When deriving (\ref{eq.cgle-ph.a},\ref{eq.cgle-ph.b}) using symmetry arguments,
one interesting aspect has to be taken into account. 
In most secondary bifurcations, the oscillating amplitude is either even or odd under reflection 
symmetry (see, for instance, \cite{CoIo90}). In our case, however,  the field ${\cal H}$ transforms under reflection 
${\bf x} \rightarrow -{\bf x}$ as ${\cal H} \rightarrow 
\overline{\cal H}$. The temporal phase of this complex amplitude is therefore
a pseudo-scalar, changing sign under reflection. This is because this phase is
related to the oscillating frequency which, in turn, depends linearly on the 
chiral symmetry breaking coefficient ($\omega_H=2\sqrt{3}R_H^2 \gamma$). This
implies that, in Eq. (\ref{eq.cgle-ph.b}), the term $({\cal H}\nabla\overline{{\cal H}} - \overline{{\cal H}}\nabla{\cal H})$ breaks the chiral symmetry, but not the term $[{\cal H}({\bf e}_z\times\nabla)
\overline{{\cal H}}-\overline{{\cal H}}({\bf e}_z\times\nabla){\cal H}]$, as
one could have naively expected. Looking at the values for the coefficients we see 
that, in fact, $\beta_2$ changes sign while $\eta$ is invariant under 
$\omega_H \rightarrow -\omega_H$.

It is interesting to note that at the band-center ($q=0$) the system (\ref{eq.cgle-ph.a},\ref{eq.cgle-ph.b}) decouples. In this case the usual 
CGLE for the amplitude of the oscillation is recovered, which after rescaling the amplitude, time, and space can be written as \cite{ChMa96}:
\begin{equation}
\partial_t H = H + (1+ib_1) \nabla^2 H - (b_3 -i\;{\rm sign}(\omega_H)) H |H|^2,
\label{eq.cgl}
\end{equation}
with
\begin{eqnarray}
&&b_1=\frac{\xi_i}{\xi_r}=
\frac{2(R_H^2+2\omega_H^2)q^2}{(2q^2R_H-R_H^2-\omega_H^2)\omega_H}, \\
&&b_3=-\frac{\rho_r}{|\rho_i |}=\frac{2|\omega_H | R_H (3R_H+1)}{\omega_H^2 (1+4R_H) 
+ 8R_H^2 (1+2R_H)},\label{eq.b3}
\end{eqnarray}
where the sub-indices r and i indicate real and imaginary part, respectively. 
Furthermore, at the band-center $b_1 = 0$. 

\begin{figure}
\centerline{
\epsfxsize=4cm\epsfbox{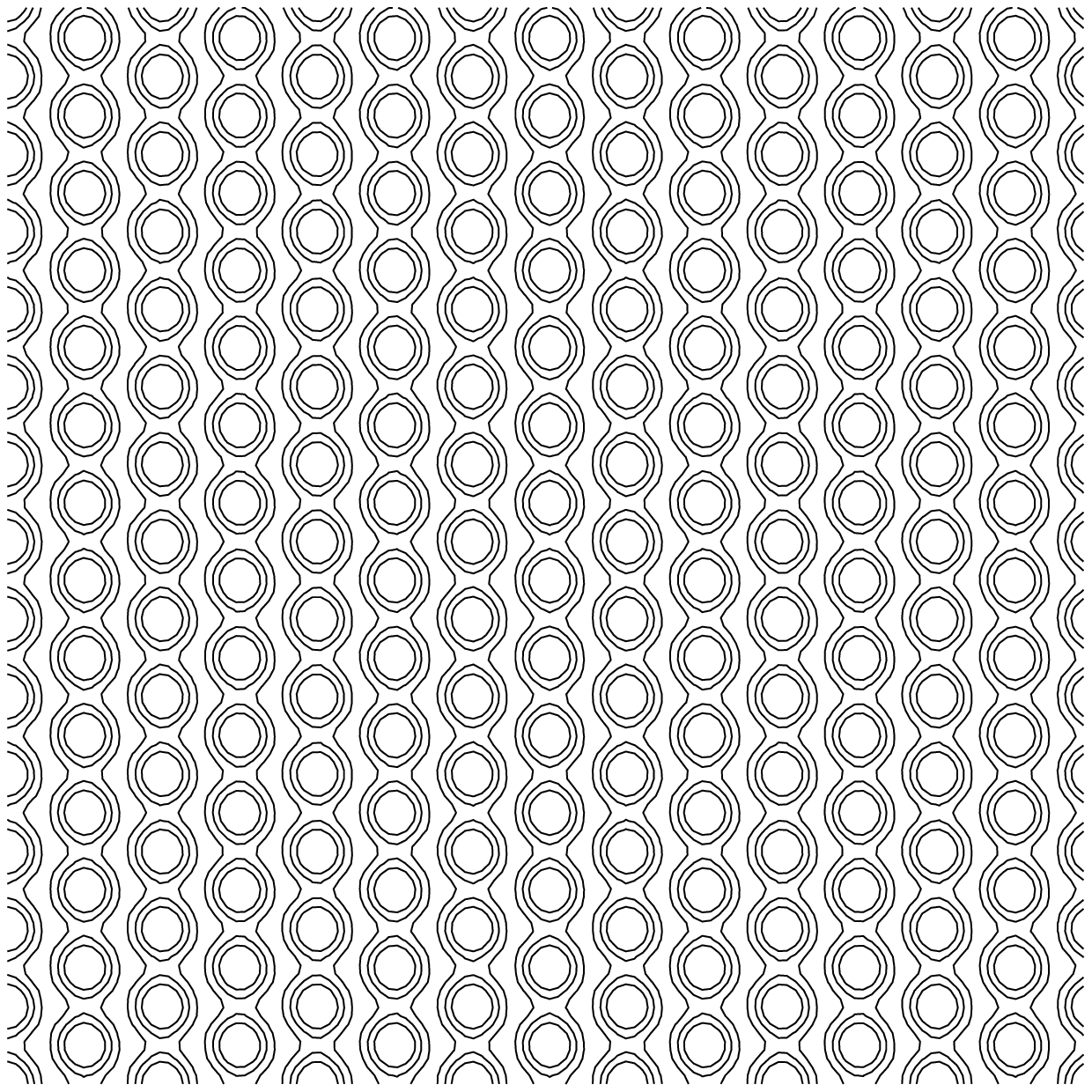}\hspace{0.5cm}
\epsfxsize=4cm\epsfbox{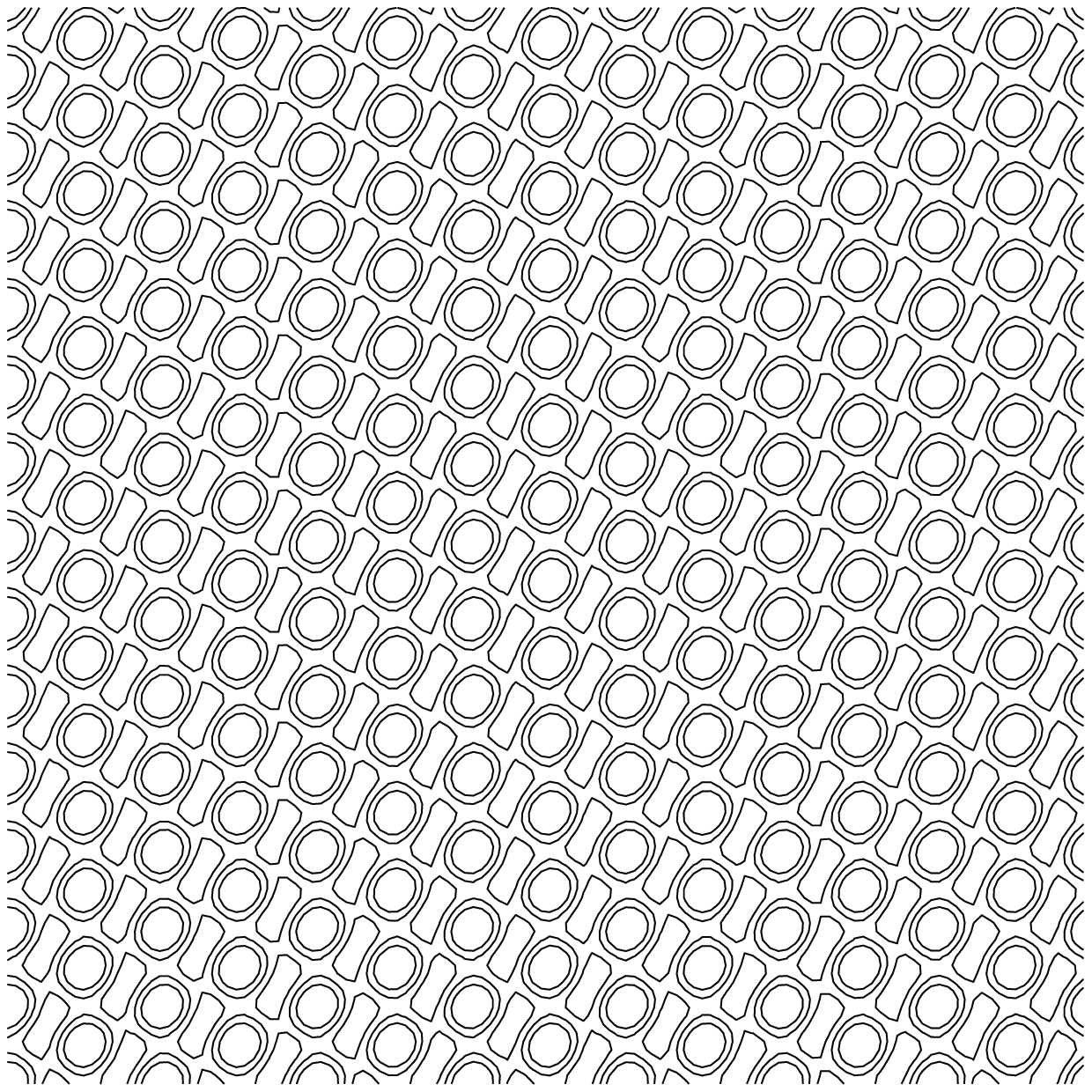}\hspace{0.5cm}
\epsfxsize=4cm\epsfbox{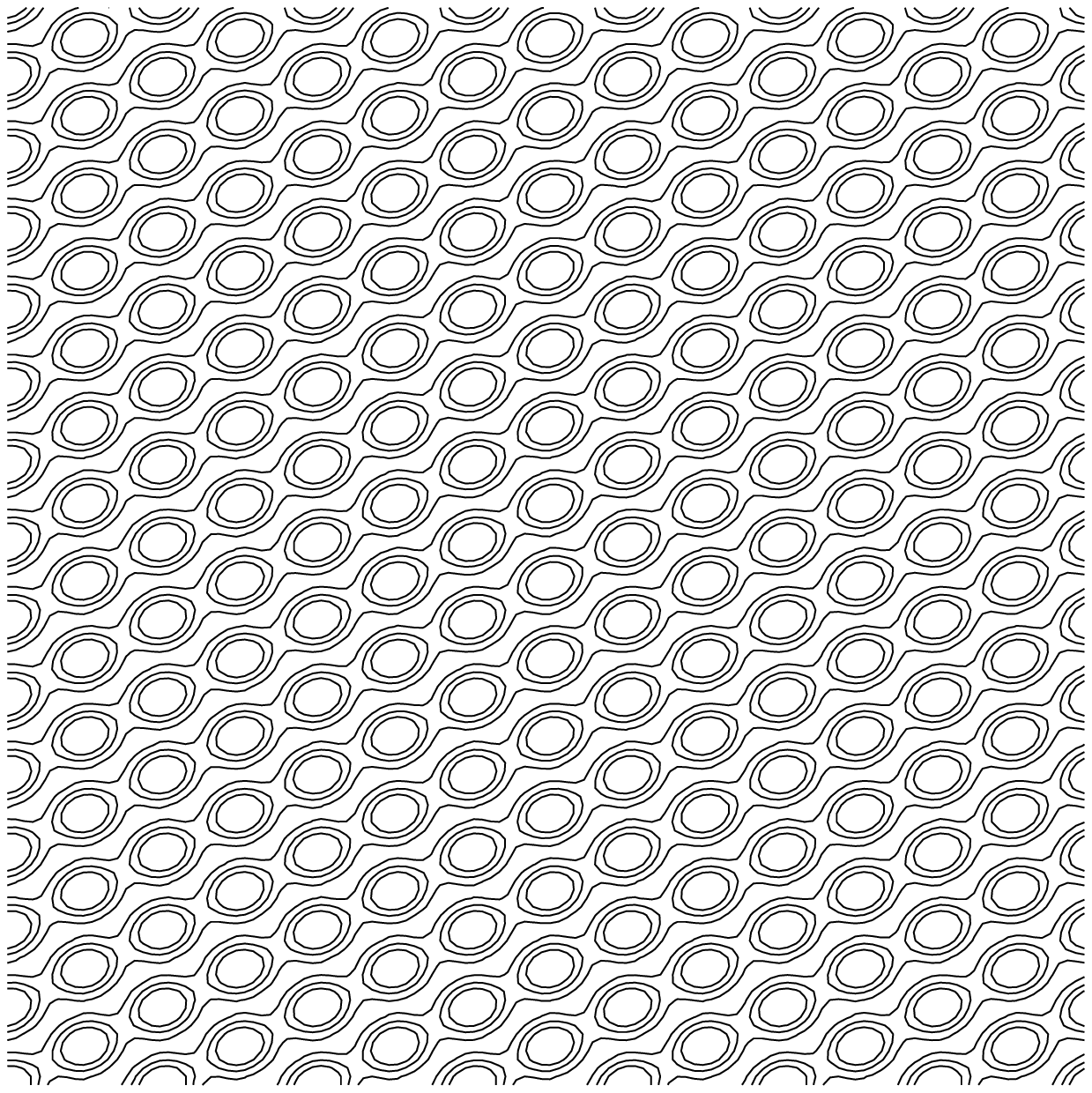}
}
\vspace{0.5cm}
\centerline{
\epsfxsize=4cm\epsfbox{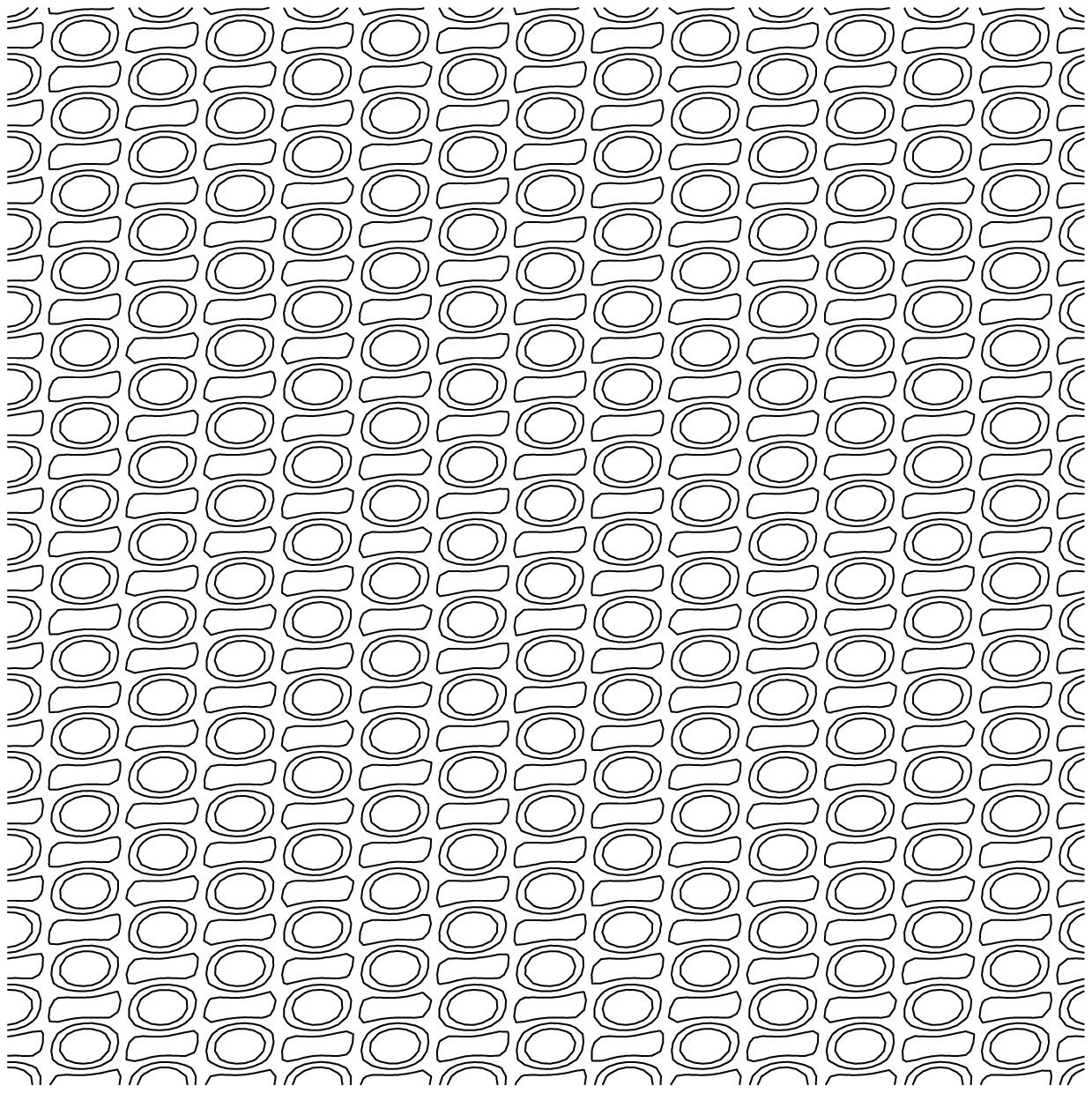}\hspace{0.5cm}
\epsfxsize=4cm\epsfbox{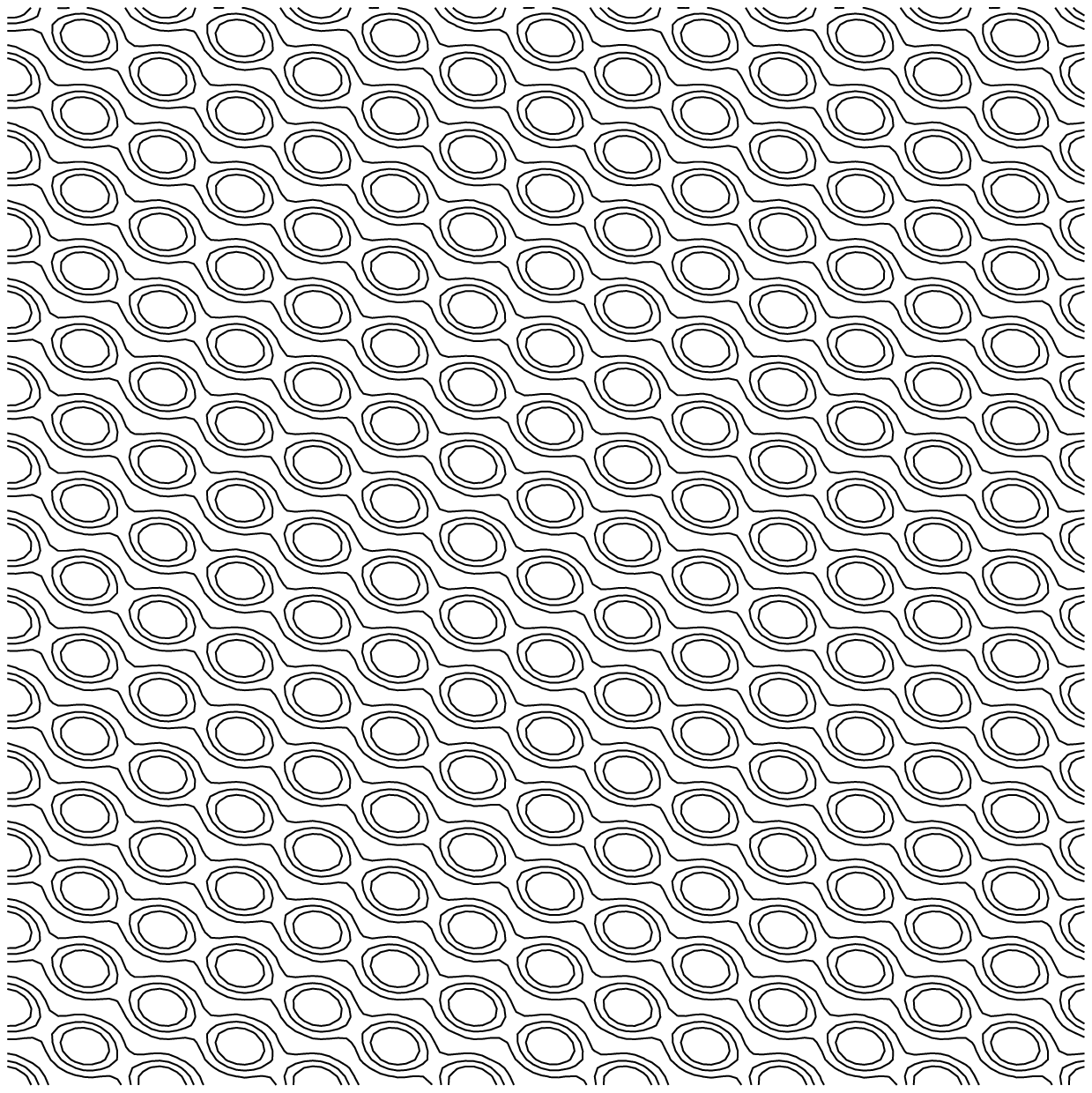}\hspace{0.5cm}
\epsfxsize=4cm\epsfbox{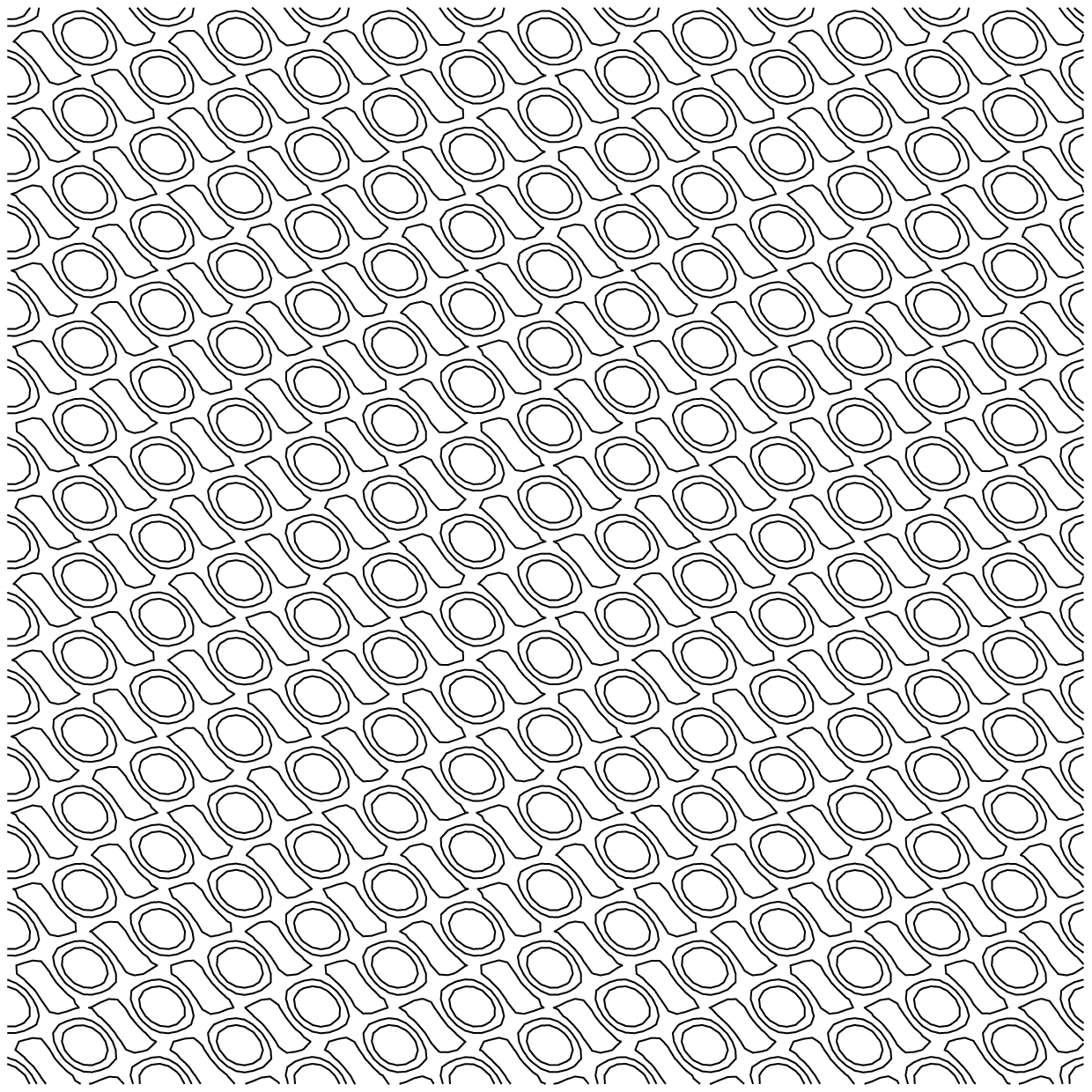}
}
\caption{Oscillating hexagons 
($\mu=4.6$, $\nu=2$, $\gamma=0.5$, $L=200$, ${\bf k}=0$, $q=0$,
$\tilde{q}_c=0.44$). There is one period of oscillation between
the first and last snapshots. The contour lines are taken at 
$\psi=\sum^{3}_{j=1} A_j e^{i\tilde{\bf q}^c_j \cdot {\bf x}} = -1,0,1$.}
\label{fig.oschex}
\end{figure}

\begin{figure}
\centerline{
\epsfxsize=4cm\epsfbox{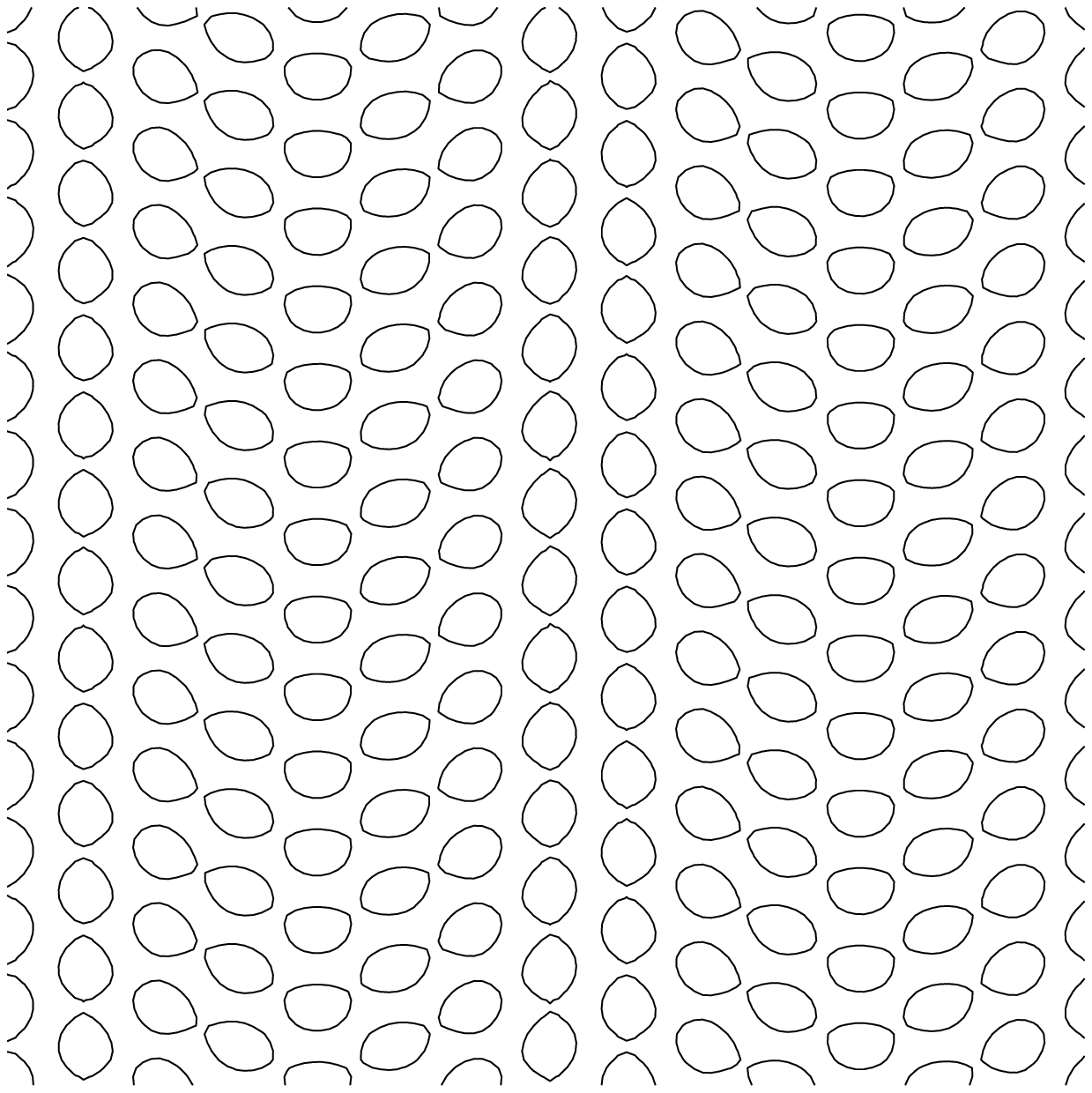}\hspace{0.5cm}\epsfxsize=4cm\epsfbox{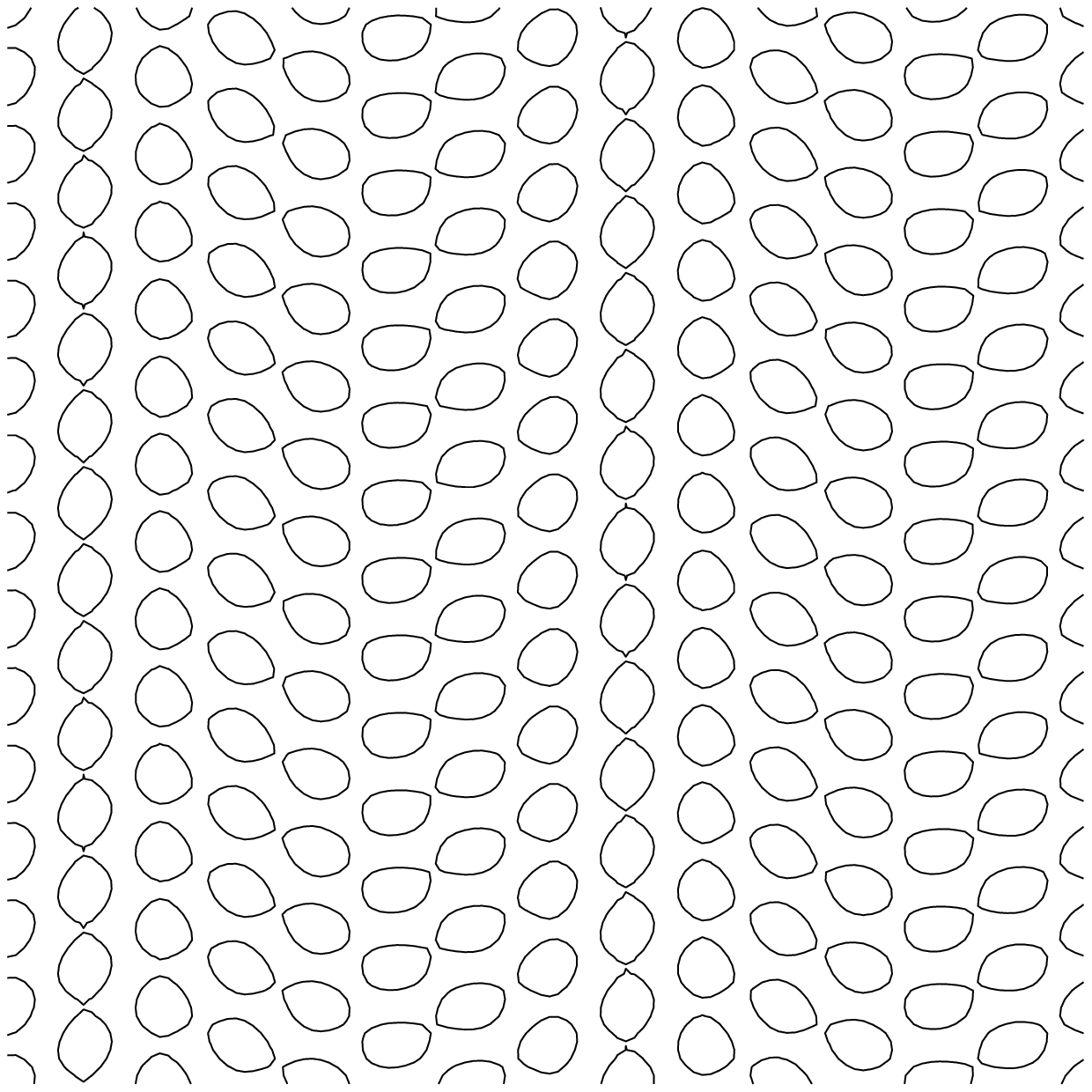}
\hspace{0.5cm}\epsfxsize=4cm\epsfbox{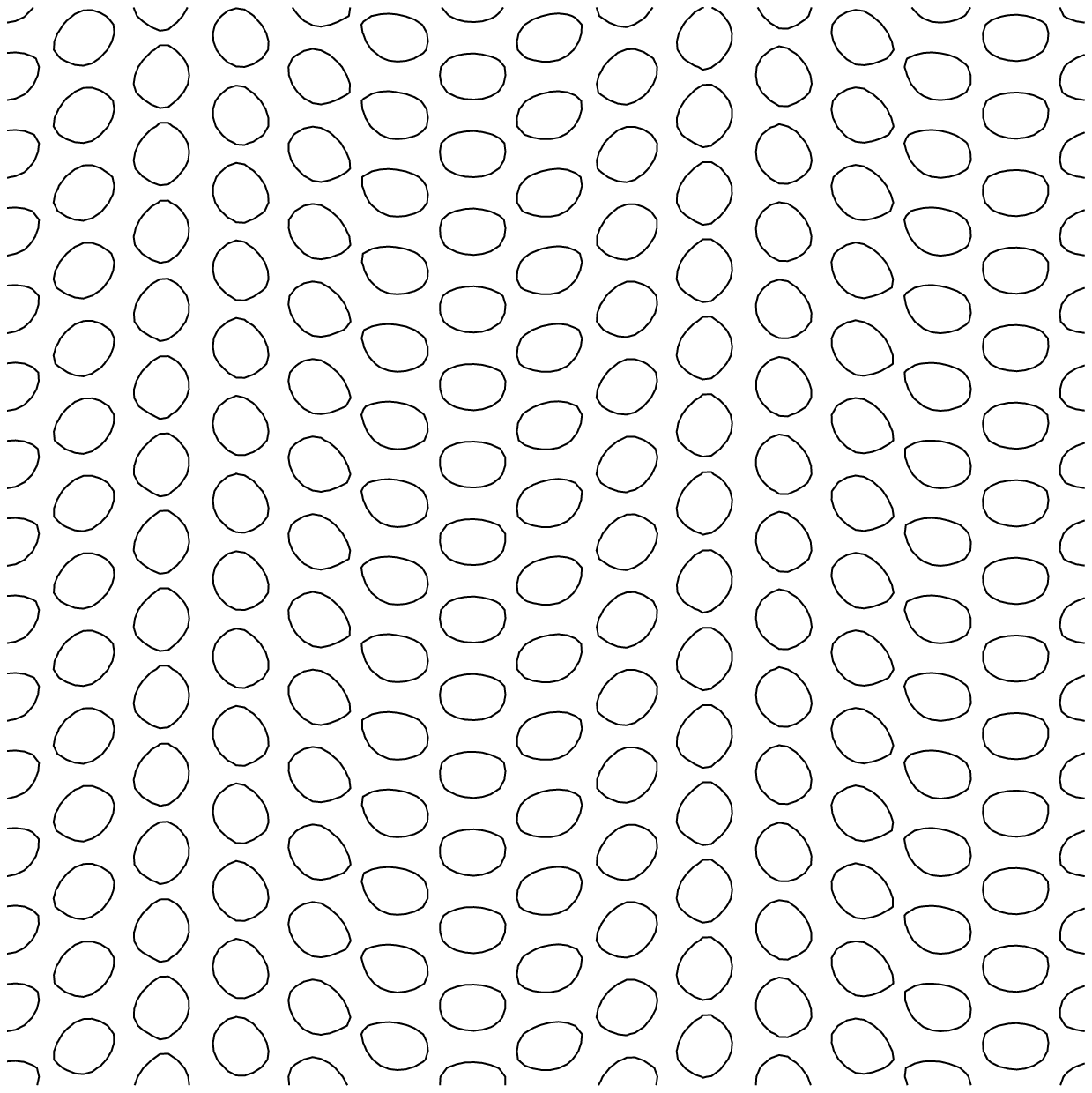}
}
\vspace{0.5cm}
\centerline{
\epsfxsize=4cm\epsfbox{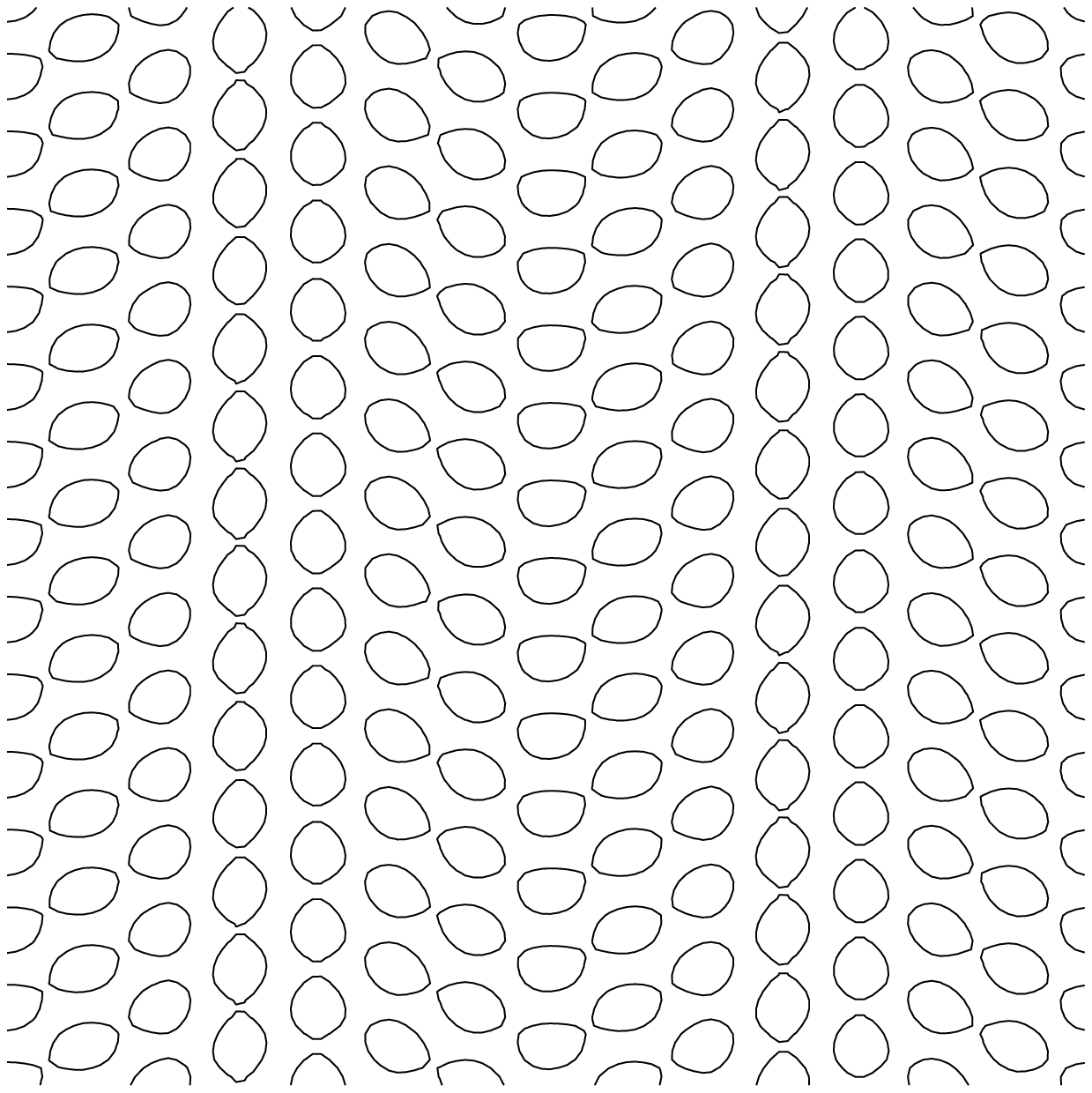}\hspace{0.5cm}\epsfxsize=4cm\epsfbox{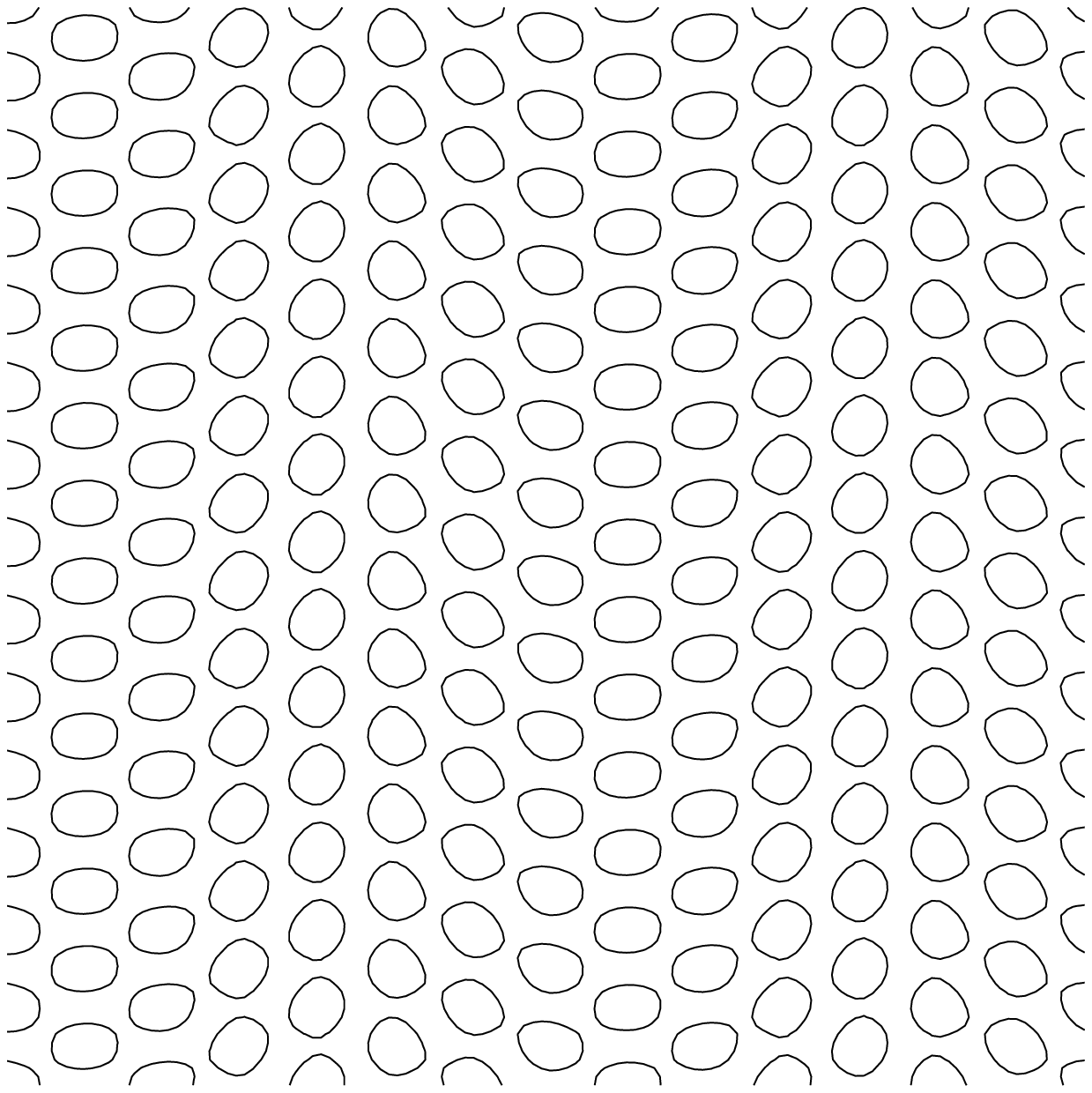}
\hspace{0.5cm}\epsfxsize=4cm\epsfbox{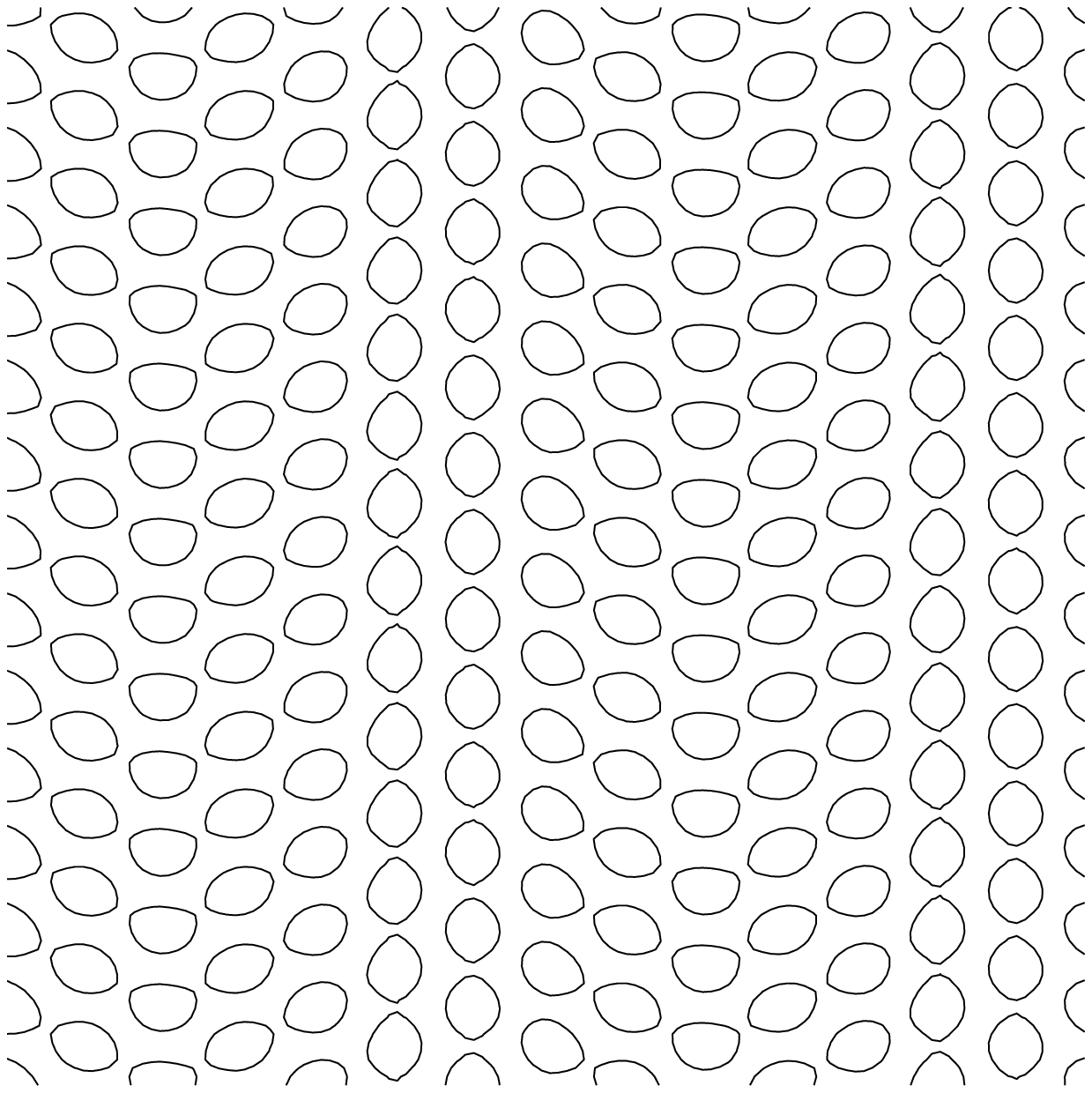}
}

\caption{Traveling hexagons ($\mu=4.6$, $\nu=2$, $\gamma=0.5$, $L=200$, $k_x=0.0628$, $k_y=0$, $q=0.125$, $\tilde{q}_c=0.31$). There is one period of oscillation between
the fist and last snapshots. The modulation is traveling with phase velocity:
$v_{ph}=(\omega_H + \Omega)/k\simeq 27.27$. The underlying hexagons are traveling with
a speed: $v=2H^2\sqrt{\beta_2^2 + \eta^2}/\tilde{q}\simeq 0.00035$, too small to be observed 
in the figures.}

\label{fig.travhex}

\end{figure}

The CGLE has been studied extensively. It possesses an
extraordinary variety of solutions, including a phase chaotic state, defect 
chaos, a frozen vortex state and stable plane waves \cite{ChMa96,MaCh96,GrJa96,MoHe96,MoHe97,To96,ToFr97,PoSt95}. For the case considered 
here the values of the parameters $b_1$ and $b_3$ are always such that the system is in a regime in which stable plane waves coexist with defect chaos. In the present context it has to be emphasized that the complex amplitude of the 
CGLE represents the amplitude of oscillation of the hexagon pattern. Thus, a solution of Eq. (\ref{eq.cgle-ph.a}) with spatially uniform amplitude ${\cal H}=He^{i\Omega t}$,
\begin{equation}
H=\sqrt{\frac{\varepsilon\delta_{1r}}
{\rho_r}}=\sqrt{\frac{\varepsilon R_H}{4(1+3R_H)}},\makebox[1cm]{}
\Omega=\delta_{1i}\varepsilon- \rho_i H^2,
\label{eq.oh}
\end{equation}
corresponds to a state in which all three amplitudes 
oscillate in time phase-shifted with respect to each other 
but the phase of each amplitude
is constant in space, as illustrated in Fig. \ref{fig.oschex}. In a traveling wave solution (TW),
\begin{equation}
{\cal H}=He^{i\Omega t+i{\bf k} \cdot {\bf x}} \label{eq.tw.1},
\end{equation}
with
\begin{equation}
H=\sqrt{\frac{\varepsilon\delta_{1r}-\xi_r k^2}{\rho_r}},\;\;\;\Omega=\varepsilon (\delta_{1i}-\frac{\rho_i}{\rho_r}\delta_{1r}) - (\xi_i -
\frac{\rho_i}{\rho_r}\xi_r)k^2,\;\;k=|{\bf k}|, \label{eq.tw.2}
\end{equation}
the phase of 
each amplitude is space dependent and in different parts of the system roll-like
hexagons 
with different orientation are dominant at any given time. This is shown in Fig. \ref{fig.travhex}. 
Note that such a state has two different wavenumbers:
that of the underlying regular hexagon pattern and that of the wave modulating the 
oscillation amplitude. At the band-center, i.e. for hexagons that have the critical 
wavenumber, the modulation of the oscillation amplitude does not affect the phase of the underlying 
hexagons.

Away from the band-center ($q\neq 0$) the complex amplitude ${\cal H}$ and the 
phase vector $\vec{\phi}$ become coupled. For the traveling-wave state of 
the oscillation amplitude (\ref{eq.tw.1}) this implies that
even the underlying hexagon pattern drifts with a velocity that 
depends on the wave vector of the
modulation. Specifically, the phase is given by
 $\vec{\phi}=\vec{\omega}t$, with
\begin{equation}
\vec{\omega}=-2H^2[\beta_2 {\bf k} - \eta (\hat{\bf e}_z \times {\bf k})],
\label{eq.omega}
\end{equation}
which implies a drift of the hexagons with a speed $v=2H^2\sqrt{\beta_2^2 +
\eta^2}/\tilde{q}$, at an angle $\theta=\arctan(-\eta/\beta_2)$ with
respect to the wave vector of the traveling wave. 
Substituting the values of the coefficients (\ref{coeffs.1}-\ref{coeffs.2}) 
one obtains for the angle
 $\theta=\arctan(3R_H/\omega_H)$. Thus, for $\omega_H\;\rightarrow 0$ 
 the angle becomes $\theta=\pm \pi/2$, i.e. the drift is perpendicular to the
wave vector of the modulation.
When  $\omega_H \rightarrow \infty$, on the other hand, they drift in the parallel direction ($\theta\; \rightarrow \; 0,\pi$). The speed $v$ is given by  
\begin{equation}
v=4\left (\frac{H}{R_H} \right )^2 \frac{qk}{\tilde{q}\sqrt{1+\left (\frac{\omega_H}{3R_H} \right )^2}}.
\end{equation}

The traveling waves exist above the curve $\varepsilon \delta_{1r} - \xi_r k^2 =0$, up to the global bifurcation at $\mu=\mu_{het}$. However they can be 
unstable to side-band perturbations.
In the following we study their stability properties. 
In particular, we will focus  on the 
effect of the coupling of the oscillation amplitude ${\cal H}$ to 
the phase of the underlying hexagons.

\section{Linear stability analysis}

 To consider the linear stability of the oscillating hexagons,
 we perturb the traveling-wave state (\ref{eq.tw.1}) as
\begin{equation}
{\cal H}=(H+h)e^{i(\Omega t + {\bf k}\cdot{\bf x} + \varphi)},
\;\;\;\vec{\phi}=\vec{\omega}t + \vec{\phi}_{1}.\label{eq.exp}
\end{equation}
For simplicity we write $\vec{\phi}_{1}$ in the following as 
$\vec{\phi}$. 
Substituting  (\ref{eq.exp}) in Eqs. (\ref{eq.cgle-ph.a},\ref{eq.cgle-ph.b}) 
we obtain the linear equations for the perturbations:

\begin{eqnarray}
\partial_T h&=&-2\xi_r H {\bf k}\cdot \nabla \varphi 
-2\xi_i {\bf k}\cdot \nabla h + \xi_r\nabla^2 h-
\xi_i H \nabla^2 \varphi-\delta_{2r} H \nabla\cdot
\vec{\phi} \label{eq.hk}\\
&& -2\rho_r H^2 h, \nonumber \\
\partial_T \varphi&=&-2\xi_i {\bf k}\cdot \nabla \varphi + 
\frac{2 \xi_r}{H} {\bf k}\cdot \nabla h + \xi_r\nabla^2\varphi+
\frac{\xi_i}{H}\nabla^2 h-\delta_{2i}
\nabla\cdot\vec{\phi}-2\rho_i H h, \label{eq.varphik}\\
\partial_T\vec{\phi}&=&D_{\bot}\nabla^2\vec{\phi}+
D_{\|}\nabla(\nabla\cdot\vec{\phi})
+D_{\times_1}\nabla^2({\bf e}_z\times\vec{\phi})+
D_{\times_2}({\bf e}_z\times\nabla)(\nabla\cdot\vec{\phi}) \label{eq.vecphik} \\
&&+2\alpha H \nabla h+2\beta_1 H 
({\bf e}_z \times \nabla) h-2\beta_2 H^2 \nabla\varphi+
2\eta H^2({\bf e}_z\times\nabla)\varphi \nonumber \\
&&-4\beta_2 H h {\bf k} + 4\eta H h (\hat{\bf e}_z \times {\bf k}). 
\nonumber
\end{eqnarray}

This leads to a $4\times 4$ linear eigenvalue
problem, which must be solved numerically. In the long-wave limit the
perturbation in the amplitude $h$ becomes slaved to the gradients of the 
phases $\varphi$ and $\vec{\phi}$. The resulting $3 \times 3$ system is,
however, still rather involved. A substantial 
simplification occurs in the case ${\bf k}=0$, i.e. 
for homogeneously oscillating hexagons. In order to gain physical insight we 
will consider this case first.

\subsection{Homogeneous Oscillations}

For homogeneously oscillating hexagons $\vec{\omega}=0$ and the perturbation equations 
(\ref{eq.hk},\ref{eq.varphik},\ref{eq.vecphik}) reduce to
\begin{eqnarray}
\partial_T h&=&\xi_r\nabla^2 h-\xi H \nabla^2 \varphi-\delta_{2r} H \nabla\cdot
\vec{\phi}-2\rho_r H^2 h,\\
\partial_T \varphi&=&\xi_r\nabla^2\varphi+\frac{\xi_i}{H}\nabla^2 h-\delta_{2i}
\nabla\cdot\vec{\phi}-2\rho_i H h,\\
\partial_T\vec{\phi}&=&D_{\bot}\nabla^2\vec{\phi}+D_{\|}\nabla(\nabla\cdot\vec{\phi})
+D_{\times_1}\nabla^2({\bf e}_z\times\vec{\phi})+D_{\times_2}({\bf e}_z\times\nabla)(\nabla\cdot\vec{\phi}) \nonumber \\
&&+2\alpha H \nabla h+2\beta_1 H 
({\bf e}_z \times \nabla) h-2\beta_2 H^2 \nabla\varphi+2\eta H^2({\bf e}_z\times\nabla)\varphi.
\end{eqnarray}

In the limit of 
long-wave perturbations the amplitude $h$ can be eliminated adiabatically
\begin{equation}
h\simeq\frac{1}{2\rho_r H^2}[-\xi_i H\nabla^2\varphi-\delta_{2r} H \nabla\cdot
\vec{\phi}].
\end{equation}
In this manner we arrive at a system for the three phases, 
two corresponding to the spatial translations in the plane, 
the other to a temporal shift,
\begin{eqnarray}
\partial_T\vec{\phi}&=&D_{\bot}\nabla^2 \vec{\phi}+\left (D_{\|}-\frac{\alpha\delta_{2r}}
{\rho_r}\right )\nabla(\nabla\cdot\vec{\phi})+D_{\times_1}\nabla^2({\bf e}_z\times
\vec{\phi}) \nonumber \\
&&+\left (D_{\times_2}-\frac{\beta_1\delta_r}{\rho_r}\right )({\bf e}_z\times\nabla)(\nabla\cdot
\vec{\phi})-\frac{\alpha\xi_i}{\rho_r}\nabla^2(\nabla\varphi) \nonumber \\
&&-\frac{\beta_1\xi_i}{\rho_r}
\nabla^2({\bf e}_z\times\nabla)\varphi-2\beta_2 H^2\nabla\varphi+2\eta H^2
({\bf e}_z\times\nabla)\varphi,\label{eq.lw1}\\
\partial_T\varphi&=&\left (\xi_r+\frac{\rho_i}{\rho_r}\xi_i\right )\nabla^2\varphi
+\left (\delta_{2r}\frac{\rho_i}{\rho_r}-\delta_{2i}\right )\nabla\cdot\vec{\phi}.\label{eq.lw2}
\end{eqnarray}

At the band-center, $q=0$, Eqs. (\ref{eq.cgle-ph.a},\ref{eq.cgle-ph.b}) 
decouple since $\beta_1=\beta_2=\delta_2=\alpha=\eta=0$ 
(cf. (\ref{coeffs.1}-\ref{coeffs.2})). It is easy to show that in 
that case the eigenvalues are
\begin{eqnarray}
&&\sigma_{1,2}=-\frac{1}{2}\left [2D_{\bot}+
D_{\|} \pm \sqrt{D_{\|}^2-4D_{\times_1} (D_{\times_1} + 
D_{\times_2})}\right ] Q^2, \label{eq.ncs12}\\
&&\sigma_3=-\left (\xi_r+\frac{\rho_i}{\rho_r}\xi_i \right ) Q^2. \label{eq.ncs3}
\end{eqnarray}
We obtain therefore the usual expression for the phase instabilities of the underlying hexagons \cite{EcRipre} and the Benjamin-Feir instability of the 
oscillations \cite{BeFe67}. The actual values
of these eigenvalues, when $q=0$, are $\sigma_1=-Q^2/4$, $\sigma_2=-3Q^2/4$ and 
$\sigma_3=-Q^2/2$. The system is therefore always stable at the band-center.

Away from the band-center ($q\ne0$) the system is no longer decoupled. 
At leading order in the long-wave expansion (\ref{eq.lw1}), (\ref{eq.lw2}) 
we have then
\begin{eqnarray}
\partial_T {\vec \phi}&=&-2\beta_2 H^2 \nabla \varphi + 
2\eta H^2 (\hat{e}_z \times \nabla) \varphi,\\
\partial_T \varphi &=& \left (\delta_{2r}\frac{\rho_i}{\rho_r}-
\delta_{2i}\right ) \nabla \cdot \vec{\phi}.\label{eq.div}
\end{eqnarray}
These two equations can be combined into a single 
second-order equation for $\vec{\phi}$,
\begin{equation}
\partial^2_{T}\vec{\phi}= 
-2 \left (\delta_{2r}\frac{\rho_i}{\rho_r}-
\delta_{2i}\right )H^2 (\beta_2 \nabla  -
\eta (\hat{e}_z\times\nabla))(\nabla \cdot \vec{\phi}).
\end{equation}

Writing in normal modes, $\phi_x=\phi^{0}_{x}e^{i{\bf Q}\cdot{\bf x}+
\sigma t}$, $\phi_y=\phi^{0}_{y}e^{i{\bf Q}\cdot{\bf x}+\sigma t}$, 
$\varphi=\varphi^0 e^{i{\bf Q}\cdot{\bf x}+
\sigma t}$, $Q\equiv |{\bf Q}|$, we arrive at the dispersion relation
\begin{eqnarray}
\sigma_1&=&0, \label{eq.s1Q}\\
\sigma_{2,3}&=&\pm\sqrt{2 
\left (\delta_{2r}\rho_i/\rho_r-\delta_{2i}\right )\beta_2} 
\, H Q, \label{eq.s23Q}
\end{eqnarray}
indicating the possibility of two different instabilities.

The eigenvalue $\sigma_1$ corresponds to the divergence-free part of 
$\vec{\phi}$ and does not involve $\varphi$ (cf. Eq. (\ref{eq.div})). It
is marginal at this order. The eigenmodes associated with $\sigma_{2,3}$
involve both phases 
$\vec{\phi}$ and $\varphi$. For $\gamma^2 > 2/(3R_H^2)$, $\sigma_2$  
 is 
always positive (except at
the band-center, where $\beta_2=0$). Hence there exists a critical value for the
rotation $|\gamma_c|\equiv \sqrt{2/3}/R_H$ above which the system is 
unstable for $Q \rightarrow 0$. When $|\gamma|< |\gamma_c|$ the eigenvalues $\sigma_{2,3}$ 
are purely imaginary and the stability is determined at next order. 

At quadratic order in the perturbation wavenumber $Q$ one 
obtains 
\begin{eqnarray}
\sigma_1&=&\left (-D_{\bot}+\frac{\eta}{\beta_2}D_{\times_1}\right ) 
Q^2,
\label{eq.s1gen}\\
\sigma_{2,3}&=&\pm \sqrt{2H^2\beta_2\left (\delta_{2r}\frac{\rho_i}{\rho_r}
-\delta_{2i}\right )}\,Q \label{eq.s23gen} \\
&&-\frac{1}{2}
\left ( D_{\bot}+D_{\|}-\alpha\frac{\delta_{2r}}{\rho_r}+\beta_2  \frac{\delta_{2r} \rho_i}{\rho_r^2}+\frac{\eta}{\beta_2}
D_{\times_1}+\xi_r+\frac{\rho_i}{\rho_r}\xi_i\right)\,Q^2. \nonumber
\end{eqnarray}

\begin{figure}

\centerline{\epsfxsize=7cm\epsfbox{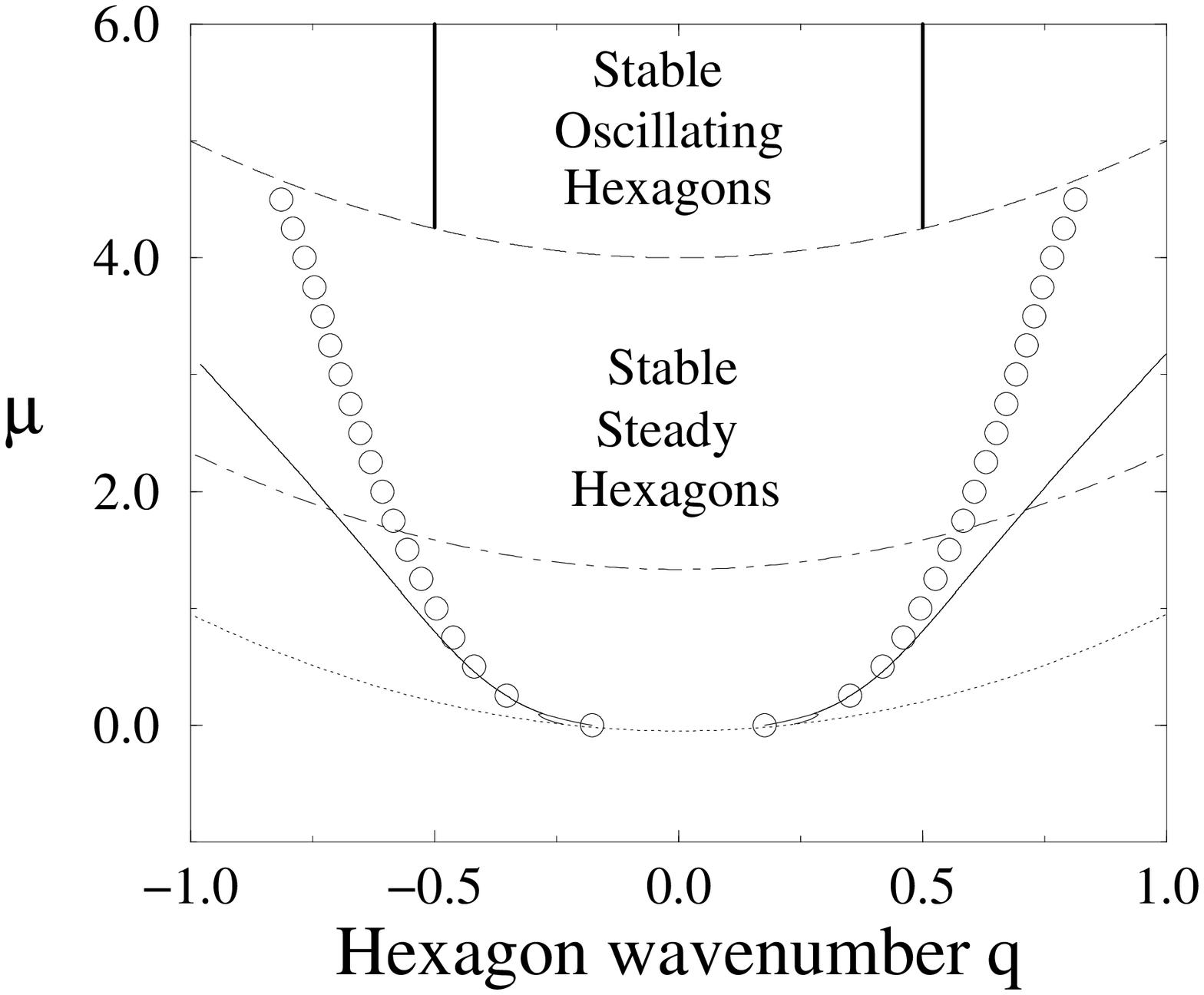}\epsfxsize=7cm\epsfbox{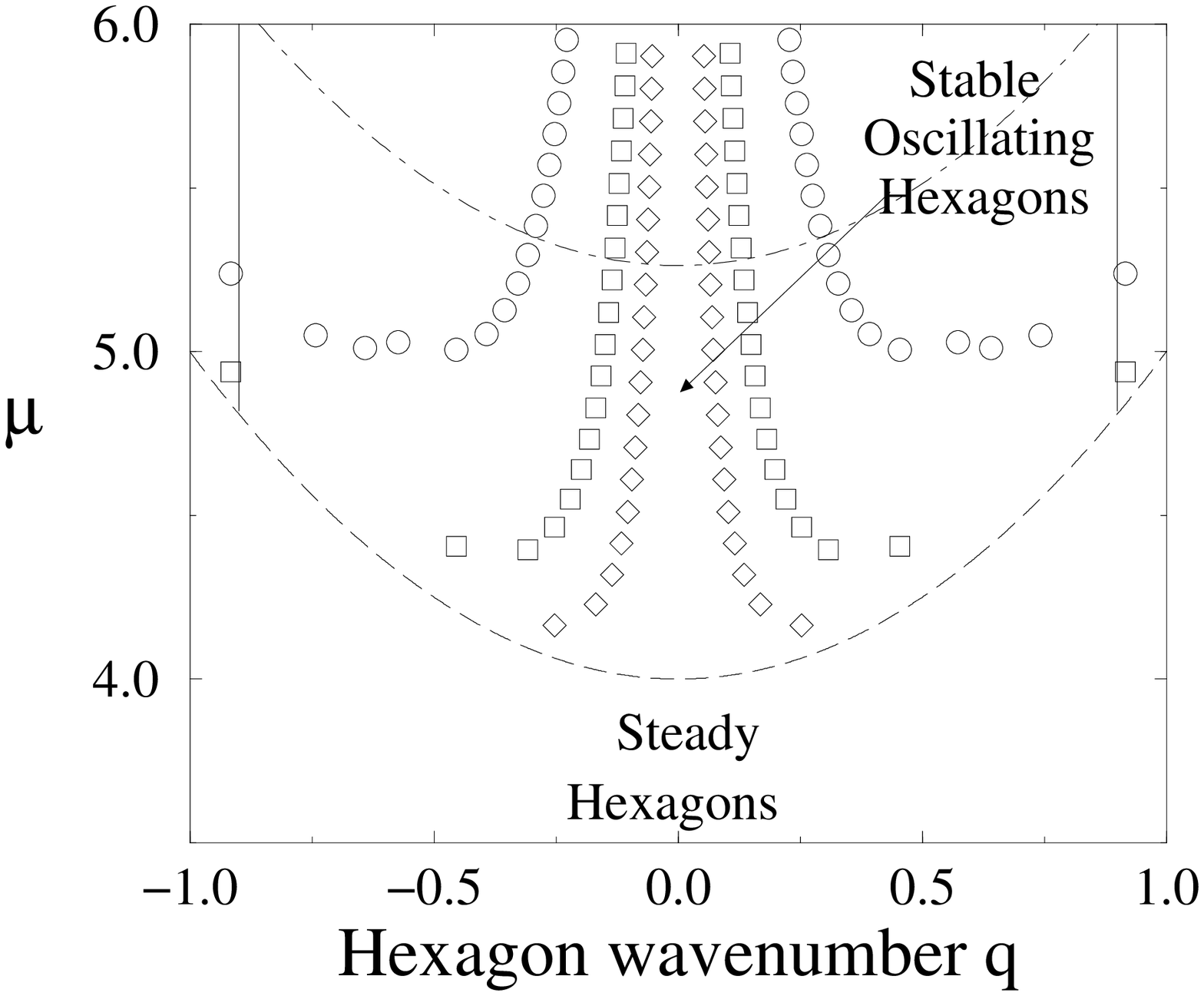}}

\caption{Stability limits of oscillating hexagons ($\nu=2$, $\gamma_c=0.816$). 
Lines indicate saddle-node for hexagons (dotted), Hopf bifurcation (dashed) 
and stability limits for rolls against hexagons (dot-dashed). a) $\gamma=0.5< 
\gamma_c$. 
b) $\gamma=0.9>\gamma_c $. Symbols indicate different system sizes, L=250 
(circles), 500 (squares) and 1000 (diamonds). The stable region close to the band-center shrinks as the system 
size is increased.} 
\label{fig.staboh}
\end{figure}

Substituting the values of the coefficients from 
(\ref{coeffs.1}-\ref{coeffs.2}) in $\sigma_1$ we see
that it becomes positive provided $q^2 > \gamma^2 R_H^3$. In Fig. 
\ref{fig.staboh}a the long-wave stability limits of the oscillating 
hexagons are shown for $\nu=2$ and $\gamma=0.5$. These results agree
with those obtained solving the full $4 \times 4$ dispersion relation. Hence, 
for any given 
value of the rotation rate there is a value of the wavenumber above which the
system becomes unstable. The range of
stable wavenumbers decreases as the rotation rate is decreased. In fact, it 
vanishes as $\gamma \rightarrow 0$. In this limit also the range in $\varepsilon$ 
over which the oscillating hexagons exist vanishes.
Also shown in Fig. 
\ref{fig.staboh}a are the instabilities of steady hexagons, below the Hopf
bifurcation. The solid line represents the long-wave results, while the circles
are the results obtained solving the $6 \times 6$ dispersion relation 
associated with Eq. (\ref{eq.amp}) \cite{EcRipre}. The dash-dotted line 
represents the line above which the rolls become stable (cf. Eq. (\ref{eq.trR})).  

As $\gamma$ is increased the range of wavenumbers that are stable with respect 
to the diffusive mode ($\sigma_1 < 0$) increases. However, for $\gamma > 
\gamma_c$, $\sigma_2$ becomes positive for $Q \rightarrow 0$. In
finite systems the term quadratic in $Q$ may not be negligible. In fact, 
inserting the values from (\ref{coeffs.1}-\ref{coeffs.2}) 
into Eq. (\ref{eq.s23gen}) shows that the $Q^{2}$-term in $\sigma_{2,3}$ is 
always stabilizing 
when $q\ll 1$ ($\sigma_2=\sigma_3=-5Q^2/8$,
when $q=0$), and for $q \sim {\cal O}(1)$ it is only destabilizing when 
$\gamma, \nu \gg 1$. For $q=1$ typical values of $\gamma, \nu$ for which this 
happens are $\nu \simeq 30,\gamma \simeq 70$. Therefore, 
in finite systems, in which $Q_{\rm min}=2\pi /L$ cannot be arbitrarily small,
 there is always a region close to the band-center that is stable, 
 even when $|\gamma|>|\gamma_c|$. The stability limit $\sigma_2 =0$ is given 
by an expression of the form $H^2 \sim \mu-\mu_H=f(q)/(Lq^2)$, where $f(q)$ 
appears due to the $q$-dependence of the second term in (\ref{eq.s23gen}). This 
situation is shown in Fig. \ref{fig.staboh}b, where the
different symbols correspond to several values
of the system size. For smaller systems the stable region increases. The shape of the stability limits suggest that $f(q)$ does 
not depend strongly on $q$. In Fig. \ref{fig.staboh}b the instability 
corresponding to $\sigma_1 >0$ has moved to higher values of $q$ 
($q= \pm 0.9$).\footnote{The form of the eigenvalues $\sigma_{2,3}$ is similar to that 
encountered in the case of a secondary Hopf bifurcation off a roll pattern 
\cite{DaLe92,Sa93}. However, in contrast to the one-dimensional case
there is a third eigenvalue, $\sigma_1$, because the phase has two components.
Similar expressions for the eigenvalues are to be expected for other 
two-dimensional patterns, e.g. squares, undergoing a Hopf bifurcation.}

It should be noted that taking the limit $q\rightarrow 0$ in 
(\ref{eq.s1gen}), (\ref{eq.s23gen}) does not give the same results 
as in (\ref{eq.ncs12}), (\ref{eq.ncs3}). In
fact, in the limit $q \rightarrow 0$, the three eigenvalues in 
(\ref{eq.s1gen}), (\ref{eq.s23gen}) become
 $\sigma_1=-Q^2/4$, $\sigma_{2}=\sigma_{3}=-5Q^2/8$, while Eqs.  
 (\ref{eq.ncs12}), (\ref{eq.ncs3}) yield $\sigma_2=-3Q^2/4$ and 
 $\sigma_3=-Q^2/2$. This difference is due to the fact that, in order to 
obtain (\ref{eq.s1gen}) and 
(\ref{eq.s23gen}) we assume $q$ to be ${\cal O}(1)$, and expand in terms of 
$Q$,
while (\ref{eq.ncs12}) and (\ref{eq.ncs3}) are obtained by taking the limit 
$q\rightarrow 0$ first. If we consider both to be small and of the same order, 
$q \sim Q \ll 1$, then:
\begin{eqnarray}
\sigma_1 & = & -D_{\bot}Q^2 \\
\sigma_{2,3} & = & -\frac{1}{2}(D_{\bot}+D_{\|}+\xi_r)Q^2 \\
&&\pm \left [2H^2\beta_2\left ( \delta_{2r}\frac{\rho_i}{\rho_r} - 
\delta_{2i}\right ) + 
\frac{1}{4}(D_{\bot}+D_{\|}-\xi_r)^2 Q^2 \right ]^{1/2} Q. \nonumber 
\end{eqnarray}
When $q=0$ we obtain $\sigma_1=-Q^2/4$, $\sigma_2=-3Q^2/4$ and 
$\sigma_3=-Q^2/2$, while considering $Q \ll q$ expressions 
(\ref{eq.s1Q}), (\ref{eq.s23Q}) are recovered at leading order.

\subsection{Stability of Traveling Waves}

For the traveling waves the expressions become quite complicated. We therefore 
present
the analytical results from the long-wave analysis only up to linear order in the gradients. As 
in the case of homogeneous oscillations, 
in the long-wave limit $h$ becomes slaved (cf. (\ref{eq.hk})),
\begin{equation}
h \simeq \frac{1}{2\rho_r H^2}[-2\xi_r H ({\bf k}\cdot \nabla )\varphi - \delta_{2r} H (\nabla \cdot \vec{\phi})].\label{eq.hslav}
\end{equation}

At the band-center $\vec{\phi}$ and $\varphi$ decouple and we recover 
expression (\ref{eq.ncs12}) for the eigenvalues $\sigma_{1,2}$ associated with 
$\vec{\phi}$. 
The eigenvalue $\sigma_3$ describes the long-wave behavior of the single CGLE.
It is given by
\begin{equation}
\sigma_3=i{\bf v}_g \cdot {\bf Q}+ 2(1+\frac{\rho_i^2}{\rho_r^2})\frac{\xi_r^2({\bf k} \cdot {\bf Q})^2}{\varepsilon\delta_{1r}-\xi_r k^2} - (\xi_r + \frac{\rho_i}{\rho_r}\xi_i)Q^2,
\end{equation}
and yields the usual Eckhaus stability limit for waves.
Here we have introduced the group velocity
\begin{equation}
{\bf v}_g \equiv \frac{\partial \Omega}{\partial {\bf k}}=-2(\xi_i -
\frac{\rho_i}{\rho_r}\xi_r){\bf k}.
\end{equation}
The expression $\varepsilon\equiv \mu - \mu_H=\xi_r k^2 /\delta_{1r}$ gives the
neutral surface for the appearance of traveling waves of wavenumber ${\bf k}$.

Away from the band-center 
Eqs. (\ref{eq.hk},\ref{eq.varphik},\ref{eq.vecphik}) reduce at leading 
order in the long-wave expansion to
\begin{eqnarray}
\partial_T {\vec \phi}&=&\frac{4\xi_r}{\rho_r}[\beta_2 {\bf k} - \eta (\hat{\bf e}_z \times {\bf k})]({\bf k}\cdot \nabla)\varphi
+ \frac{2\delta_{2r}}{\rho_r}[\beta_2 {\bf k} - \eta (\hat{\bf e}_z \times {\bf k})](\nabla \cdot \vec{\phi}) \\
&& - 2H^2[\beta_2  \nabla \varphi - \eta  (\hat{e}_z \times \nabla) \varphi], 
\nonumber\\
\partial_T \varphi &=& -2(\xi_i -\frac{\rho_i}{\rho_r}\xi_r) {\bf k}\cdot \nabla \varphi + \left (\delta_{2r}\frac{\rho_i}{\rho_r}-\delta_{2i}\right ) \nabla \cdot \vec{\phi}.
\end{eqnarray}

This yields the eigenvalues 
\begin{eqnarray}
\sigma_1&=&0\\
\sigma_{2,3}&=&\frac{i}{2}\left ({\bf v}_g + \frac{\delta_{2r}}{\rho_r}{\bf v}_h \right ) \cdot {\bf Q} \pm \left \{ -\frac{1}{4}\left [\left ({\bf v}_g + \frac{\delta_{2r}}{\rho_r}{\bf v}_h \right )\cdot {\bf Q}\right ]^2 \right .\\
&&\left . + \frac{4}{\rho_r} (\delta_{2r} {\bf v}_g \cdot {\bf Q} - \xi_r \Gamma {\bf k}
\cdot {\bf Q})({\bf v}_g \cdot {\bf Q}) + 2H^2 \beta_2 \Gamma Q^2 \right \}^{1/2},\nonumber
\end{eqnarray}
where we have defined $\Gamma\equiv \delta_{2r}\rho_i/\rho_r - \delta_{2i}$ 
and ${\bf v}_h\equiv -\vec{\omega}/(2H^2)$ (cf. (\ref{eq.omega})).

\begin{figure}

\centerline{
\epsfxsize=7cm\epsfbox{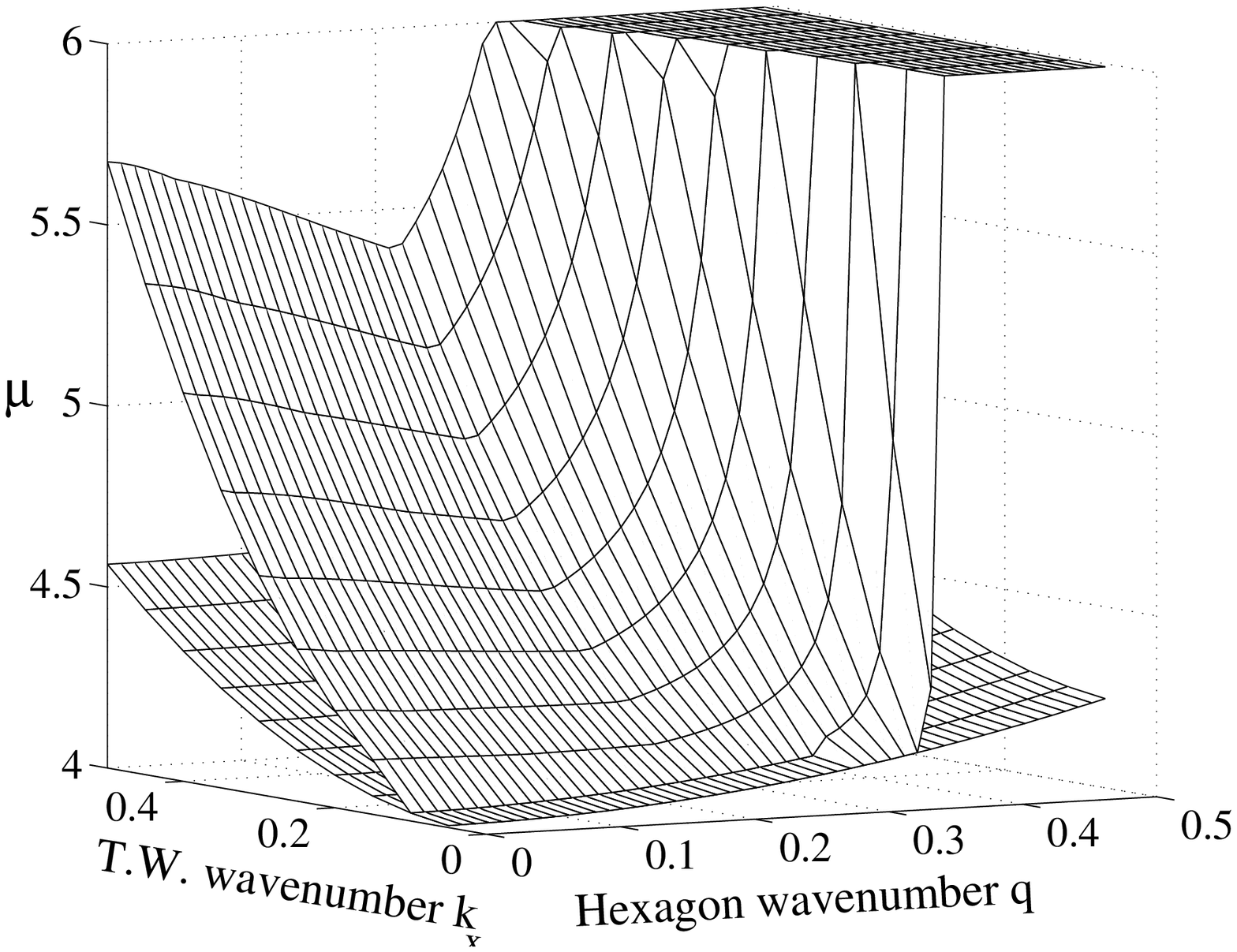}
\epsfxsize=7cm\epsfbox{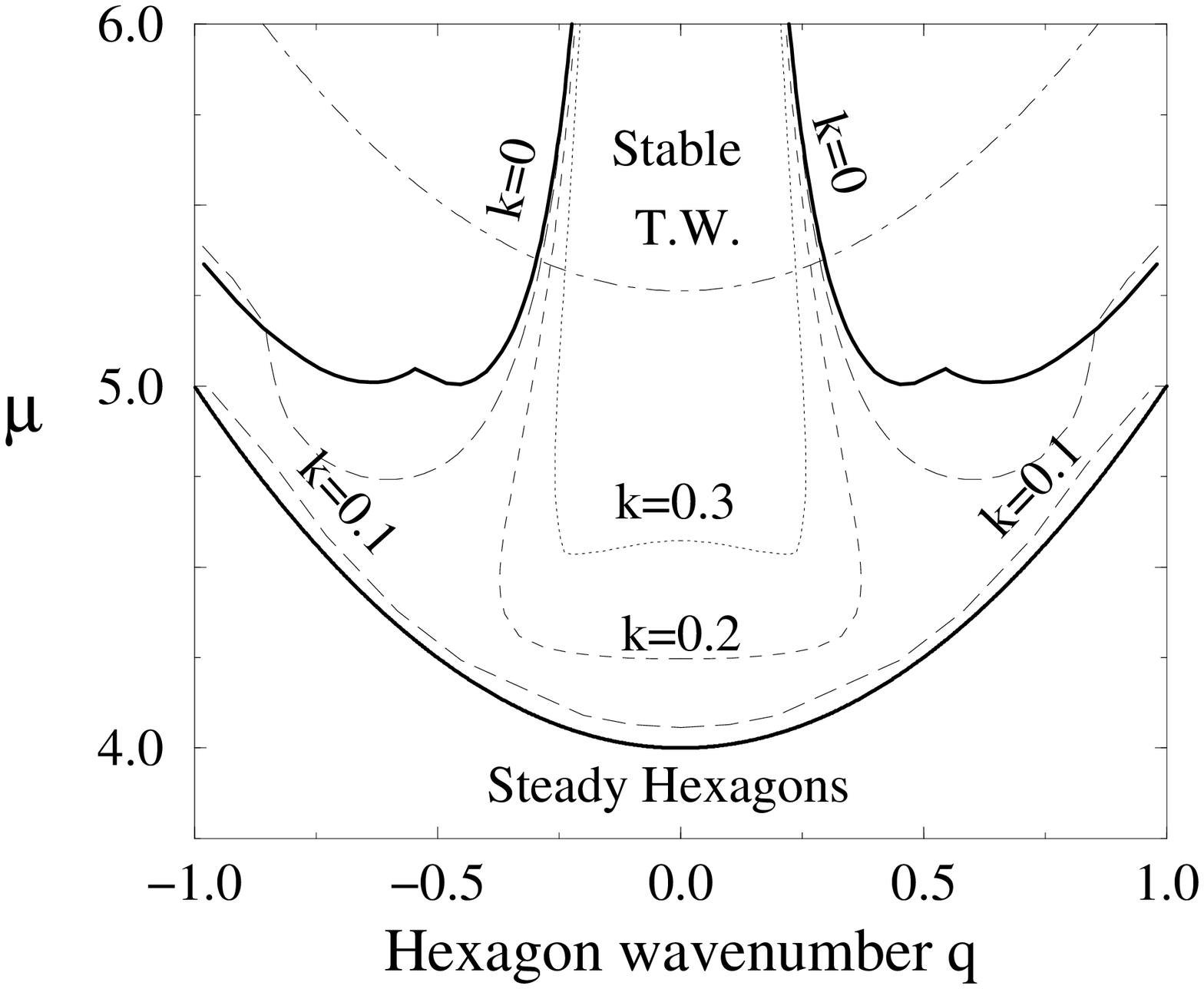}
}

\caption{Stability limits of traveling waves 
($\nu=2$, $\gamma_c=0.816$). a) $\gamma=0.3<\gamma_c$. Shown are the neutral 
surface and the stability limit. Traveling waves are stable above the second 
surface. b) $\gamma=0.9>\gamma_c$. Dashed-dotted line: stability limit of 
rolls against hexagons.
} 
\label{fig.stabohk}
\end{figure}

Since the analytical expressions for the eigenvalues are quite complicated  
we solve Eqs. (\ref{eq.hk},\ref{eq.varphik},\ref{eq.vecphik}) directly without 
the additional long-wave approximation (\ref{eq.hslav})
for various values of the 
wavenumber ${\bf k}=(k,0)$ of the traveling waves. 
 Fig. \ref{fig.stabohk}a shows the neutral surface of
 traveling waves and their stability limit as a function of the 
hexagon wavenumber $q$. Due to the 
 reflection symmetries $q \rightarrow -q$ and $k \rightarrow -k$  
 only one quadrant is shown. For clarity the stability surface has 
 been capped at $\mu=6$. For $k=0$ the stability limit does not
 depend on $\mu$ and the stability surface is vertical (cf. Fig. 
\ref{fig.staboh}a). 
 As $k$ is increased the stability surface becomes smoother. 
For $k \neq 0$, the waves are unstable at onset and become stable above
the second surface.  
 Fig. \ref{fig.stabohk}b shows cross-sections of the stability 
 surface for $\gamma > \gamma_{c}$, for a system of size $L=250$. When $k=0$
we recover the results from Fig. \ref{fig.staboh}b. For $k\neq 0$, but small 
(cf. $k=0.1$ in Fig. \ref{fig.stabohk}b) the traveling waves are unstable 
at onset but they become stable for larger values of $\mu$. When $|q|$ is 
large they can become unstable again as $\mu$ is increased. For larger values 
of $k$, the latter two stability lines merge and the stability region becomes 
bounded in $q$. At this point the stability limits in Figs. 
\ref{fig.stabohk}a and \ref{fig.stabohk}b look similar.

\section{Numerical simulations}

In order to study the nonlinear behavior arising from the 
instabilities, we have performed numerical simulations of Eqs.  
(\ref{eq.amp}) and (\ref{eq.cgle-ph.a}), (\ref{eq.cgle-ph.b}).  
A Runge-Kutta method with an integrating factor 
that computes the linear derivative terms exactly has been used.  
Derivatives were computed in Fourier space, using a two-dimensional 
fast Fourier transform (FFT).  The numerical simulations were done in 
a rectangular box of aspect ratio $2/\sqrt{3}$ with periodic boundary 
conditions.  This aspect ratio was used to allow for regular hexagonal 
patterns.

\begin{figure}

\centerline{\epsfysize=5.5cm\epsfbox{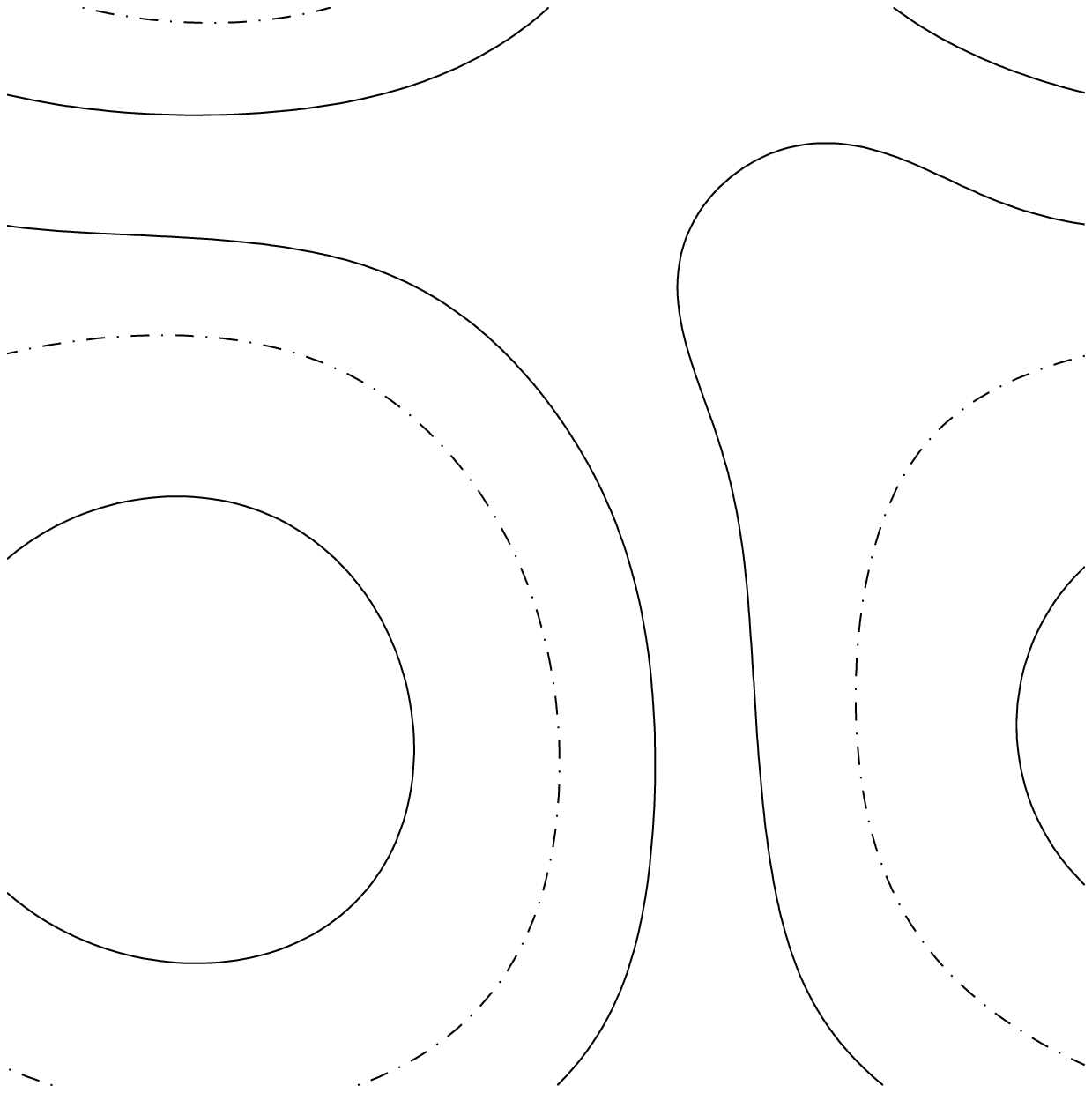}\hspace{2cm}\epsfysize=5.5cm\epsfbox{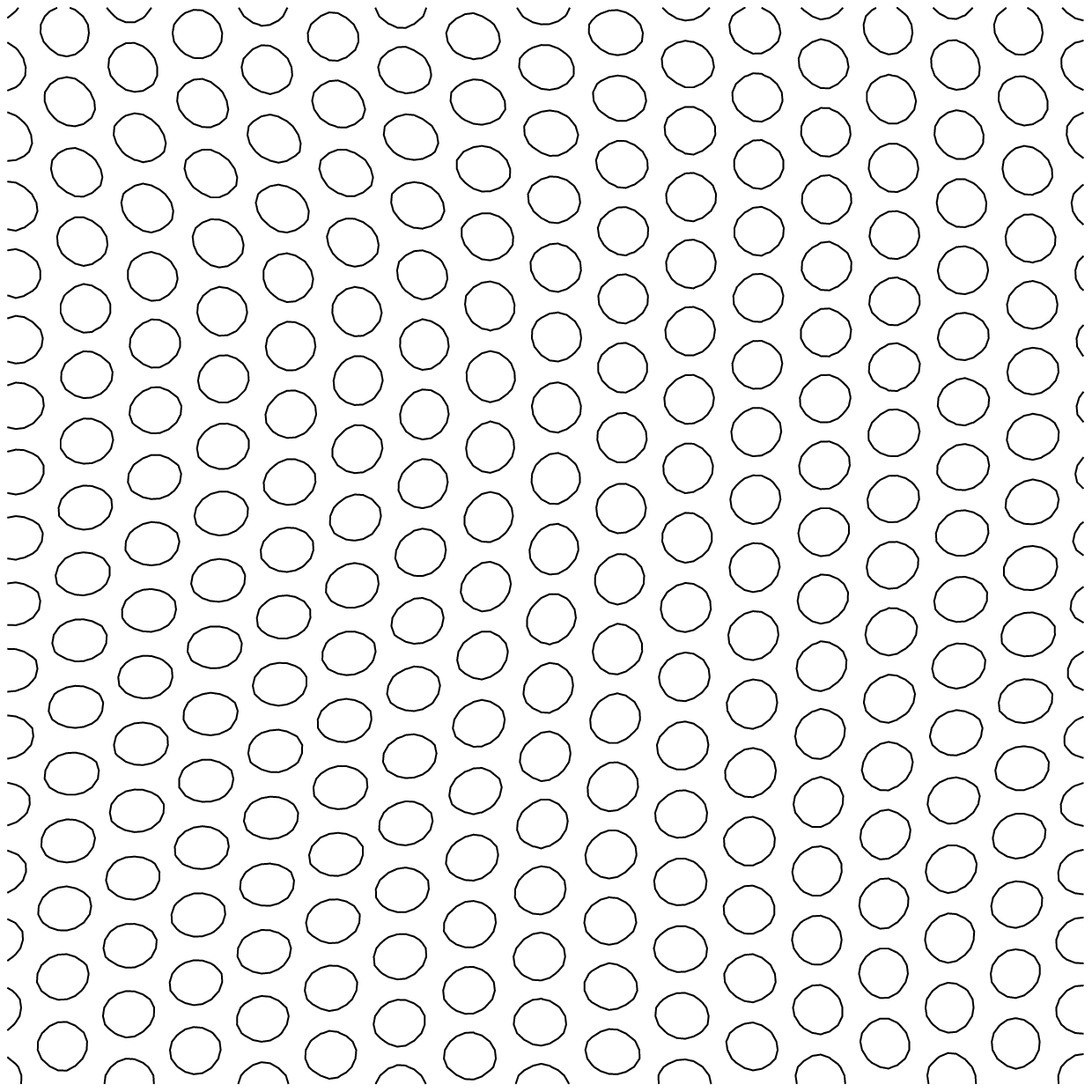}}

\caption{State resulting from the long-wave instability $\sigma_1 > 0$ 
($\nu=2$, $\gamma=0.3$, $q=0.5$, $\mu-\mu_H=0.1$, $L=250$). a) The dashed and 
solid lines mark the zero contour lines of the real and 
imaginary part of the oscillation amplitude ${\cal H}$. b) Reconstruction of 
the hexagonal pattern from a), with $\tilde{q}_c=0.4$}
\label{fig.num}
\end{figure}

We investigate the stability of oscillating hexagons simulating both the 
original amplitude equations (\ref{eq.amp}) and the reduced amplitude-phase 
equations (\ref{eq.cgle-ph.a}), (\ref{eq.cgle-ph.b}). To that end, we start 
with homogeneously oscillating hexagons as given by Eq. (\ref{eq.oh}) (or Eq. 
(\ref{eq.expan})) and add weak noise. For values of $\mu$ close to $\mu_H$, the 
growth rates obtained from (\ref{eq.amp}) and  (\ref{eq.cgle-ph.a}), 
(\ref{eq.cgle-ph.b}) agree with each other and with the results from the linear
stability analysis. Both the long-wave instabilities coming from $\sigma_1 > 0$
(\ref{eq.s1gen}) and $\sigma_{2,3} > 0$ (\ref{eq.s23gen}) lead to 
qualitatively similar behavior. The perturbations grow until they 
reach a saturation suggesting that the bifurcations are supercritical.
Since the perturbations involve spatial modulations of the oscillation 
amplitude the hexagons begin to travel. However, 
the modulation wavevector ${\mathbf{k}}$ varies in space and 
induces drift velocities that are different in magnitude and direction 
at different locations in the system, implying a shear of the pattern. This 
results in deformed hexagon patterns as shown in Fig. \ref{fig.num}.

\begin{figure}

\centerline{\epsfxsize=8cm\epsfbox{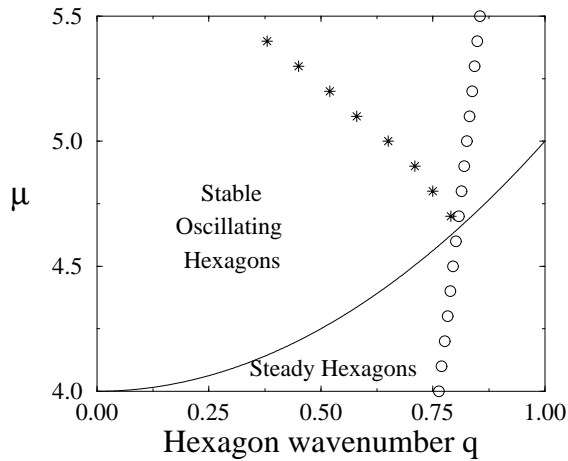}}
\caption{Short-wave stability of oscillating hexagons (stars) and steady
hexagons (circles) for the same parameters as in Fig. \ref{fig.staboh}a 
($\nu=2$, $\gamma=0.5$). Solid
line: Hopf bifurcation of steady hexagons.}
\label{fig.stabsw}

\end{figure}

In addition to the long-wave instabilities a short-wave instability appears for
larger values of the hexagon wavenumber $q$. This instability is 
induced by the short-wave 
instability of the steady hexagons \cite{EcRipre} 
and cannot be studied with the amplitude-phase equations (\ref{eq.cgle-ph.a}), (\ref{eq.cgle-ph.b}). 
As $\mu$ is increased above 
$\mu_H$, the oscillating hexagons arise through a Hopf bifurcation off 
the steady
hexagons. Since that bifurcation occurs at zero wavenumber it affects the 
long-wave properties of the system and we expect that 
the long-wave stability limits of the steady and of the oscillating 
hexagons differ qualitatively. The effect of the Hopf bifurcation 
on short-wave instabilities, on the other hand, should vanish
as the amplitude of the oscillations goes to 0. Thus, 
we expect a continuous transition from the 
short-wave instability of the steady to that of the oscillating hexagons
as the line $\mu=\mu_H$ is crossed. In order to check this, we 
numerically determine the short-wave 
instability of the oscillating hexagons and compare it with the stability results 
for the steady hexagons, as obtained from solving the sixth-order dispersion
relation associated with Eq. (\ref{eq.amp}) \cite{EcRipre}. The transition is 
indeed continuous, as expected. The stability region of the oscillating
hexagons turns out to be reduced as compared to that of the steady hexagons when $\mu$ is 
increased (see Fig. \ref{fig.stabsw}). 

If the noise added to the oscillating hexagons is sufficiently large, 
different dynamics may arise even in the parameter range in which the 
oscillating hexagons are linearly stable. As indicated earlier, within 
the framework of the three coupled Ginzburg-Landau equations (\ref{eq.amp}) 
one obtains for the parameters in the CGLE (\ref{eq.cgl}) values for 
which a persistent chaotic state exists while the plane waves are 
linearly stable. A detailed study of the chaotic state, which 
is characterized by the creation and annihilation of defects, 
comparing its description using 
the coupled Ginzburg-Landau equations (\ref{eq.amp}), 
the amplitude-phase equations (\ref{eq.cgle-ph.a},\ref{eq.cgle-ph.b}), 
and the single CGLE (\ref{eq.cgl}) has been presented elsewhere 
\cite{EcRiunpub}. One of the main results of that study is the 
observation that this system is one of the few in which the defect 
chaos of the CGLE should be accessible experimentally.

\section{Conclusions}

In this paper we have studied the effect of a breaking of the chiral symmetry
on systems that exhibit hexagon patterns. Classic examples of such systems 
are non-Boussinesq and surface-tension-driven convection with rotation. We
have focused on the dynamics of the oscillating hexagons that arise in a Hopf
bifurcation due to the rotation. In the vicinity of this secondary Hopf
bifurcation the oscillating hexagons are described by a single complex 
Ginzburg-Landau equation (CGLE) coupled to the two phases of the underlying
hexagons. The resulting amplitude-phase equations have certain similarities
with those describing the secondary Hopf bifurcation observed in rectangle
patterns in electro-convection in nematics \cite{JaKo92}. 

Like the CGLE, the amplitude-phase equations support homogeneously oscillating 
solutions
as well as traveling waves. In the latter, the coupling to the phases induces
a drift of the hexagons in a direction that is typically oblique to the 
propagation direction of the traveling waves. The stability analysis of the
oscillating and the traveling hexagons reveal two types of long-wave
instabilities, one 
occurring when the hexagons are far away from the band-center, the other for 
high enough rotation rate. Even in this latter case, in finite systems there 
is always a stable region close 
to the band-center, its size depending on the size of the system. In both 
cases, the instabilities appear to be 
supercritical, giving rise to a spatially modulated oscillating hexagonal 
pattern. 

Although there is always a region in which the homogeneously oscillating
hexagons are linearly stable within the three coupled Ginzburg-Landau
equations, they are in fact only meta-stable. Sufficiently large perturbations
induce a transition to a state of defect chaos described by the CGLE 
\cite{EcRiunpub}. There is always bistability between the chaotic state and
the ordered oscillations. In that respect the chaotic state resembles 
spiral-defect chaos as it is observed in Rayleigh-B\'enard convection at low
Prandtl numbers \cite{CaEg97}. As in that system the ordered states can presumably only be 
obtained by carefully controlled initial conditions. Rotating hexagon 
convection appears to be the first system in which the defect chaotic regime 
of the CGLE should be accessible experimentally.

From numerical simulations 
of the amplitude equations (\ref{eq.amp}) it has been shown that a transition 
between defect-chaos and a frozen vortex state occurs for wavenumbers far 
away from the band-center \cite{EcRiunpub}. 
This transition happens when the asymptotic plane waves emitted by the 
spirals become absolutely unstable. Therefore,
the stability results for the traveling wave solution are relevant in order 
to determine the range
of existence of the chaotic state found in \cite{EcRiunpub}. A complete
treatment of this point would involve the determination of the asymptotic
wavenumber $k_\infty$ for the waves emitted by the spiral solutions of the 
coupled amplitude-phase equations.

In the present work, we have studied the system close to the bifurcation 
point, where
analytical results can be obtained. However, far away from the Hopf 
bifurcation, a number of interesting effects are expected. One is related 
to the strength of the chiral symmetry breaking. To leading order in the 
perturbation expansion, the usual single CGLE is
obtained, which is chirally symmetric. The breaking of the chiral symmetry 
appears only through the coupling with the phase, and vanishes at the 
band-center.
For larger values of the oscillation amplitude,
higher-order terms breaking the chiral symmetry of the CGLE are expected. This 
leads to an
asymmetry between defects with opposite topological charge \cite{NaOt98}, 
with one type of spirals becoming dominant in the frozen vortex state.
The asymptotic wavenumber, as well as the onset of absolute instability, 
becomes different for positively and negatively charged spirals, and the 
transition to the defect chaotic regime is different than in the chirally
symmetric case \cite{NaOt98}. We expect 
that this effect of the chiral symmetry breaking
can be observed with Eqs. (\ref{eq.amp}), as the control parameter is increased
above the Hopf bifurcation.

Another open question is the relation of the limit cycle corresponding
to the oscillating hexagons with the heteroclinic orbit arising in the 
K\"uppers-Lortz instability of rolls. As the limit cycle approaches the mixed-mode solutions, the harmonic oscillations are transformed into a switching 
between the three roll modes making up the hexagons similar to the dynamics 
arising in the K\"uppers-Lortz regime. This suggest a connection 
between the domain chaos found in these systems and the regular and disordered
states discussed in this paper. However, it is worth emphasizing that in
the defect chaos regime described in \cite{EcRiunpub}, the orientation of 
the 
hexagons is well defined and what is spatially chaotic is the modulation of
the amplitudes that compose the hexagon pattern. In the K\"uppers-Lortz regime,
on the other hand, patches of rolls with arbitrary 
orientation are possible, resulting in a state with an isotropic Fourier 
spectrum. Thus,
a quantitative comparison between both states is not possible with 
Ginzburg-Landau equations such as
(\ref{eq.amp}) and generalized Swift-Hohenberg models must be considered 
\cite{SaRi99}.

{\bf Acknowledgments}

We gratefully acknowledge interesting discussions with F. Sain and M. Silber.
The numerical simulations were performed with a modification of a code by G.D. Granzow.
This work was supported by D.O.E. Grant DE-FG02-G2ER14303 and NASA Grant NAG3-2113.

\appendix

\section{Derivation of the Amplitude-Phase Equations}

At the Hopf bifurcation, the hexagon solution (\ref{eq.hexa}) becomes 
$R_H=1/(\nu -1)$. We will consider perturbations around this solution, both in 
amplitude and phase:
\begin{equation}
A_i=(R_H+r_i)e^{i{\bf q}_i \cdot {\bf x} +\phi_i + \Phi},
\end{equation}
where $\Phi=\phi_1+\phi_2+\phi_3$ is the global phase of the hexagons and the 
three phases $\phi_i$ can be written as
\begin{eqnarray*}
\phi_1 & = & \phi_x,\\
\phi_2 & = & -\phi_x/2 +\sqrt{3}\phi_y/2,\\
\phi_3 & = & -\phi_x/2 -\sqrt{3}\phi_y/2,
\end{eqnarray*}
where we have defined the phase vector $\vec{\phi}=(\phi_x,\phi_y)$, with
$\phi_x$ and $\phi_y$ related to translations of the pattern in the x- and
y-directions.

For the perturbations of the modulus $r_i$ there are three eigenvalues, one
real, $\sigma_1=-2R^2 (1+2\nu) + R$, corresponding to an eigenvector with 
$r_1=r_2=r_3$ and a complex conjugate pair, $\sigma_{2,3}=-2R^2 (1-\nu) -2R \pm 
2\sqrt{3}R^2 \gamma i$, whose real part vanishes at $\mu=\mu_H$. The 
corresponding eigenvector satisfies $r_3=e^{2\pi i/3} r_2 = e^{4\pi i/3} r_1$.

Taking this into account, we consider the expansion:
\begin{eqnarray*}
r_1&=&\epsilon r + \left [ (-\frac{1}{2} + i\frac{\sqrt{3}}{2}) (\sqrt{\epsilon}{\cal H}e^{i\omega t} + \epsilon ({\cal H}_{10} + \overline{\cal H}_{12} e^{-2i\omega t})) + c.c.\right ],\\
r_2&=&\epsilon r + \left [ (-\frac{1}{2} - i\frac{\sqrt{3}}{2}) (\sqrt{\epsilon}{\cal H}e^{i\omega t} + \epsilon ({\cal H}_{10} + \overline{\cal H}_{12} e^{-2i\omega t})) + c.c.\right ], \\
r_3&=&\epsilon r + \left [ (\sqrt{\epsilon}{\cal H}e^{i\omega t}+\epsilon ({\cal H}_{10} + \overline{\cal H}_{12} e^{-2i\omega t})) + c.c.\right ],\\
\phi_x &=& \sqrt{\epsilon}\phi_x + \epsilon (\phi_{x1} e^{i\omega t} + c.c.),\\
\phi_y &=& \sqrt{\epsilon}\phi_y + \epsilon (\phi_{y1} e^{i\omega t} + c.c.),\\
\Phi & = & \epsilon (\Phi_1 e^{i\omega t} + c.c.).
\end{eqnarray*}

We also assume that the resulting state evolves on long time and space
scales, specifically: $\partial_t \sim {\cal O}(\epsilon)$, $\nabla \sim
{\cal O}(\epsilon^{1/2})$.
Substituting the former expressions into Eq. (\ref{eq.amp}), at ${\cal O}(\epsilon^{1/2})$ the linear problem is recovered, giving the 
value for the critical frequency: $\omega_H=-2\sqrt{3}R_H^2\gamma$. 

At ${\cal O}(\epsilon)$ an algebraic relation between the slaved and the 
marginal modes is obtained:

\begin{eqnarray}
\phi_{x1}&=&\frac{q}{2R_H}(\sqrt{3} + i)(\partial_x {\cal H} - i \partial_y {\cal H}),\\
\phi_{y1}&=&-\frac{qi}{2R_H}(\sqrt{3} + i)(\partial_x {\cal H} - i \partial_y {\cal H}),\\
\Phi_1&=&\frac{q}{2R_H(3R_H+i\omega_H)}(\sqrt{3}i - 1)(\partial_x {\cal H} + i \partial_y {\cal H}),\\
r&=&\frac{1}{3R_H(1+2R_H)}\left [\mu R_H -(1+6R_H)|{\cal H}|^2 -qR_H \nabla \cdot \vec{\phi} \right ], \\
{\cal H}_{10}&=&-\frac{qR_H}{4\omega_H}(\sqrt{3}i + 1) \left [\partial_x \phi_y
+ \partial_y \phi_x + i(\partial_x \phi_x - \partial_y \phi_y) \right ], \\
{\cal H}_{12}&=&\frac{1}{6} \left (1 + \frac{8i}{\omega_H} \right ) {\cal H}^2.
\end{eqnarray}

At ${\cal O}(\epsilon^{3/2})$ we obtain a solvability condition for ${\cal H}$
and $\vec{\phi}$:

\begin{eqnarray}
\partial_t {\cal H}&=& \mu {\cal H} + \frac{1}{2}\nabla^2 {\cal H} -q{\cal H}\nabla \cdot \vec{\phi} - \left (1+6R_H - \frac{2\omega_H i}{R_H}\right )r{\cal H} 
\nonumber \\
&&- \frac{1}{R_H^2}\left [3R_H(1+2R_H) - \frac{\omega i}{2}\right ] |{\cal H}|^2 {\cal H} + \left ( 8 - \frac{\omega i}{R_H}\right ) \overline{\cal H} {\cal H}_{12} \nonumber \\
&&+\frac{R_Hq}{4} (1+\sqrt{3}i) [\partial_x \phi_{x1} - \partial_y \phi_{y1} +
i(\partial_x \phi_{y1} + \partial_y \phi_{x1}) ] \nonumber \\
&&  + \frac{R_Hq}{2} (\sqrt{3}i+1)(\partial_x \Phi_1 - i \partial_y \Phi_1),\\
\partial_t \phi_x &=& \frac{1}{4}\nabla^2 \phi_x + \frac{1}{2}\partial_x (\nabla
\cdot \vec{\phi}) + \frac{2q}{R_H}\partial_x r \nonumber \\
&&+ \frac{q}{2R_H}\left [ (\sqrt{3}i-1)(\partial_x {\cal H}_{10} - i \partial_y {\cal H}_{10}) + c.c. \right ]\nonumber \\
&&+\frac{\omega}{4R_H}\left [ (i-\sqrt{3})\overline{\cal H}(\phi_{x1}+i\phi_{y1})
+ c.c. \right ] \nonumber \\
&&-\frac{1}{2R_H}\left [(1+\sqrt{3}i)(3R_H-\omega i) \overline{\cal H}\Phi_{1} + c.c. \right ], \\
\partial_t \phi_y &=& \frac{1}{4}\nabla^2 \phi_y + \frac{1}{2}\partial_y (\nabla
\cdot \vec{\phi}) + \frac{2q}{R_H}\partial_y r \nonumber \\
&&+ \frac{q}{2R_H}\left [ (\sqrt{3}+i)(\partial_x {\cal H}_{10} - i \partial_y {\cal H}_{10}) + c.c. \right ]\nonumber \\
&&-\frac{\omega}{4R_H}\left [ (1+\sqrt{3}i)\overline{\cal H}(\phi_{x1}+i\phi_{y1})
+ c.c. \right ]\nonumber \\
&&-\frac{1}{2R_H}\left [(\sqrt{3}-i)(3R_H-\omega i) \overline{\cal H}\Phi_{1} + c.c. \right ].
\end{eqnarray}
Substituting the expressions of the slaved modes into the former equations we obtain the amplitude-phase equations 
(\ref{eq.cgle-ph.a}), (\ref{eq.cgle-ph.b}).

\section{Nonlinear Gradient Terms}

If we retain in the Ginzburg-Landau equations (\ref{eq.amp}) the nonlinear 
gradient terms that express the dependence of the quadratic coupling term on 
the hexagon wavenumber \cite{EcRipre} we obtain
\begin{eqnarray}
\partial _{t}A_{1} & = & \mu A_{1}+({\bf {n}}_{1}\cdot \nabla 
)^{2}A_{1}+\overline{A}_{2}\overline{A}_{3} - A_{1}|A_{1}|^{2}\nonumber\\
& &-(\nu 
+\gamma )A_{1}|A_{2}|^{2}-(\nu -\gamma 
)A_{1}|A_{3}|^{2}\nonumber \\
& & + i(\alpha _{1}+\tilde{\alpha} )\overline{A}_{2}({\bf 
{n}}_{3}\cdot \nabla )\overline{A}_{3}+i(\alpha _{1}-\tilde{\alpha} 
)\overline{A}_{3}({\bf {n}}_{2}\cdot \nabla )\overline{A}_{2}\nonumber 
\\
& & + i\alpha _{2}\left(\overline{A}_{2}(\mbox{\boldmath $\tau $}_{3}\cdot 
\nabla )\overline{A}_{3}-\overline{A}_{3}(\mbox{\boldmath $\tau 
$}_{2}\cdot \nabla )\overline{A}_{2}\right). \label{eq.ampgen}
\end{eqnarray}
For the amplitude-phase equations (\ref{eq.cgle-ph.a}), (\ref{eq.cgle-ph.b}) 
we obtain then the coefficients:
\begin{eqnarray*}
&&\alpha=1+2q\alpha_1,\\
&&R_H=\frac{\alpha}{\nu-1},\\
&&\epsilon_c=\frac{\alpha^2 (2+\nu)}{(\nu-1)^2}+q^2=R_H(3R_H+\alpha)+q^2,\\
&&v=3R_H(\alpha+2R_H),\\
&&\omega_H = 2\sqrt{3}\gamma R_H^2, \\
&&\delta_1 = \frac{2\alpha R_H}{v} -\frac{2i\omega_H}{v},\\
&&\delta_2 = \frac{2\alpha R_Hq}{v} + \frac{4R_H^2 \alpha_1 (3R_H+\alpha)}{v} - \frac{2i\omega_H (q-R_H\alpha_1)}{v},\\
&&\xi = \frac{1}{2} - \frac{R_H(R_H\alpha_1 + q)}{9R_H^2 \alpha^2 + \omega_H^2}[\frac{\sqrt{3}}{2}\tilde{\alpha}\omega_H  + 3\alpha(q +\frac{R_H}{2}(\alpha_1 - \sqrt{3}\alpha_2))] - \frac{\sqrt{3}R_Hq\tilde{\alpha}}{\omega_H} \\
&&\hspace{0.7cm}- \frac{i(R_H\alpha_1 + q)}{9R_H^2\alpha^2 + \omega_H^2}\left [ \omega_H q + \frac{R_H}{2}[\omega_H (\alpha_1 -\sqrt{3} \alpha_2) - 3\sqrt{3}R_H\alpha\tilde{\alpha}] \right ] - \frac{iq^2}{\omega_H}\\
&&\hspace{0.7cm} + \frac{iR_H^2}{4\omega_H}\left [(\alpha_1 +\sqrt{3}\alpha_2 )^2 + 3\tilde{\alpha}^2 \right ]\\
&&\rho = \frac{8\alpha(3R_H+\alpha)}{v}- \frac{4i\omega_H (\alpha+4R_H)}{R_Hv} -\frac{32i\alpha^2}{3\omega_H},\\
&&D_{\bot}=\frac{1}{4} - \frac{\sqrt{3}R_Hq\tilde{\alpha}}{\omega_H},\\
&&D_{\|}=\frac{1}{2} - \frac{R_H\alpha_1 - q}{v}\left [ R_H(\alpha_1 - \sqrt{3}\alpha_2 ) -2q \right ],\\
&&D_{\times_1} = \frac{q^2}{\omega_H} - \frac{R_H^2}{4\omega_H}\left [ (\alpha_1 + \sqrt{3}\alpha_2)^2 + 3\tilde{\alpha}^2 \right ],\\
&&D_{\times_2} = \frac{\sqrt{3}R_H\tilde{\alpha}}{v}(R_H\alpha_1 - q),\\
&&\alpha=-\frac{\alpha+6R_H}{R_Hv}[2q + R_H(\sqrt{3}\alpha_2-\alpha_1)] + \frac{18\alpha^2R_H^3\alpha_1 - 2 \omega_H^2q}{R_H^2(9R_H^2\alpha^2 + \omega_H^2)},\\
&&\beta_1=\frac{6\omega_H \alpha (q +R_H\alpha_1)}{R_H(9R_H^2\alpha^2+\omega_H^2)} 
+\frac{\sqrt{3}\tilde{\alpha}}{v}(\alpha + 6R_H),\\
&&\beta_2=\frac{6\omega_H \alpha (q +R_H\alpha_1)}{R_H(9R_H^2\alpha^2+\omega_H^2)} + \frac{\sqrt{3}\tilde{\alpha}}{R_H},\\
&&\eta=\frac{[9R_H^2\alpha^2 (2q-R_H(\alpha_1 + \sqrt{3}\alpha_2)) - R_H\omega_H^2 (3\alpha_1 + \sqrt{3}\alpha_2 )]}{R_H^2(9R_H^2 \alpha^2 + \omega_H^2)} .
\end{eqnarray*}

\subsection{Comparison with the CGLE}

After rescaling we obtain the values for the coefficients of the CGLE 
(\ref{eq.cgl}). The coefficient $b_3$ now becomes:
\begin{equation}
b_3=\frac{2|\omega_H | R_H \alpha(\alpha +3R_H)}{\omega_H^2 (\alpha+4R_H) + 
8R_H^2 \alpha^2 (\alpha +2R_H)}.
\end{equation}
For $q\neq 0$ the maximum value of $b_3$ is:
\begin{equation}
b^{max}_3=\frac{\alpha + 3R_H}{\sqrt{8(\alpha + 2R_H)(\alpha + 4R_H)}},
\end{equation}
and the limits for small and large $R_H$ are the same as in (\ref{eq.b3}), 
even if now $b_3$ depends on $q$ (through $\alpha=1 + 2q\alpha_1$). At the 
band-center $b_3$ becomes the same as (\ref{eq.b3}). 

The expression for 
$b_1$ is quite involved. At the band-center it reduces to:
\begin{equation}
b_1=\frac{(9R_H^2 + \omega_H^2)[(\alpha_1+\sqrt{3}\alpha_2)^2 + 
3\tilde{\alpha}^2 ]+2\omega_H \alpha_1 [ \omega_H (\sqrt{3} \alpha_2 - 
\alpha_1) + 3\sqrt{3}R_H \tilde{\alpha}]}{2\omega_H (9R_H^2 + \omega_H^2 -
\alpha_1 R_H^2 [ 3R_H (\alpha_1 - \sqrt{3} \alpha_2 )+\sqrt{3}\tilde{\alpha} 
\omega_H])}.
\end{equation}
An important change occurs with respect to the decoupling of the 
amplitude-phase equation. For $\alpha_1 \neq 0$ the amplitude and the phases do
not decouple anymore.
It would require $\delta_{2r}=0$, $\delta_{2i}=0$. From the latter we obtain 
$q=R_H\alpha_1$, while the former implies
\begin{equation}
q=\frac{1}{4\alpha_1}\left [-(1+4R_H\alpha_1^2) \pm 
\sqrt{(1-4R_H\alpha_1)^2-48R_H^2\alpha_1^2}\right ].
\end{equation}
The only real solution for both conditions is $\alpha_1= 0$.


\begin{thebibliography}{10}

\bibitem{KuLo69}
G. K\"uppers and D. Lortz, J. Fluid Mech. {\bf 35},  609  (1969).

\bibitem{Ku70}
G. K\"uppers, Phys. Lett. A {\bf 32},  7  (1970).

\bibitem{BuHe80}
F. Busse and K. Heikes, Science {\bf 208},  173  (1980).

\bibitem{ZhEc91}
F. Zhong, R. Ecke, and V. Steinberg, Physica D {\bf 51},  596  (1991).

\bibitem{ZhEc92}
F. Zhong and R. Ecke, Chaos {\bf 2},  163  (1992).

\bibitem{NiEc93}
L. Ning and R. Ecke, Phys.~Rev. E {\bf 47},  3326  (1993).

\bibitem{HuEc95}
Y. Hu, R. Ecke, and G. Ahlers, Phys. Rev. Lett. {\bf 74},  5040  (1995).

\bibitem{HuPe98}
Y. Hu, W. Pesch, G. Ahlers, and R. Ecke, Phys. Rev. E {\bf 58},  5821  (1998).

\bibitem{XiGu94}
H. Xi, J. Gunton, and G. Markish, Physica A {\bf 204},  741  (1994).

\bibitem{TuCr92}
Y. Tu and M. Cross, Phys. Rev. Lett. {\bf 69},  2515  (1992).

\bibitem{FaFr92}
M. Fantz, R. Friedrich, M. Bestehorn, and H. Haken, Physica D {\bf 61},  147
  (1992).

\bibitem{NeFr93}
M. Neufeld, R. Friedrich, and H. Haken, Z. Phys. B {\bf 92},  243  (1993).

\bibitem{ClKn93}
T. Clune and E. Knobloch, Phys. Rev. E {\bf 47},  2536  (1993).

\bibitem{CrMe94}
M. Cross, D. Meiron, and Y. Tu, Chaos {\bf 4},  607  (1994).

\bibitem{PoPa97}
Y. Ponty, T. Passot, and P. Sulem, Phys. Fluids {\bf 9},  67  (1997).

\bibitem{Sw84}
J. Swift,  in {\em Contemporary Mathematics Vol. 28} (American Mathematical
  Society, Providence, 1984), p.\ 435.

\bibitem{So85}
A. Soward, Physica D {\bf 14},  227  (1985).

\bibitem{MiPe92}
J. Mill\'an-Rodr\'{\i}guez {\it et~al.}, Phys. Rev. A {\bf 46},  4729  (1992).

\bibitem{EcRiunpub}
B. Echebarria and H. Riecke, preprint, {\tt patt-sol/9912002}.  .

\bibitem{CoIo90}
P. Coullet and G. Iooss, Phys.~Rev.~Lett. {\bf 64},  866  (1990).

\bibitem{DaLe92}
F. Daviaud {\it et~al.}, Physica D {\bf 55},  287  (1992).

\bibitem{Sa93}
H. Sakaguchi, Prog. Theor. Phys. {\bf 89},  1123  (1993).

\bibitem{JaKo92}
B. Janiaud, H. Kokubo, and M. Sano, Phys. Rev. E {\bf 47},  2237  (1993).

\bibitem{Ch61}
S. Chandrasekhar, {\em Hydrodynamic and Hydromagnetic Stability} (Clarendon,
  Oxford, 1961).

\bibitem{EcRipre}
B. Echebarria and H. Riecke, to appear in Physica D, {\tt patt-sol/9906008}  .

\bibitem{LaMe93}
J. Lauzeral, S. Metens, and D. Walgraef, Europhys. Lett. {\bf 24},  707
  (1993).

\bibitem{Ho95}
R. Hoyle, Appl. Math. Lett. {\bf 9},  81  (1995).

\bibitem{ChMa96}
H. Chat\'{e} and P. Manneville, Physica A {\bf 224},  348  (1996).

\bibitem{MaCh96}
P. Manneville and H. Chat\'e, Physica D {\bf 96},  30  (1996).

\bibitem{GrJa96}
G. Grinstein, C. Jayaprakash, and R. Pandit, Physica D {\bf 90},  96  (1996).

\bibitem{MoHe96}
R. Montagne, E. Hern\'andez-Garc\'{\i}a, and M.~S. Miguel, Physica D {\bf 96},
  47  (1996).

\bibitem{MoHe97}
R. Montagne, E. Hern\'andez-Garc\'{\i}a, A. Amengual, and M.~S. Miguel, Phys.
  Rev. E {\bf 56},  151  (1997).

\bibitem{To96}
A. Torcini, Phys. Rev. Lett. {\bf 77},  1047  (1996).

\bibitem{ToFr97}
A. Torcini, H. Frauenkron, and P. Grassberger, Phys. Rev. E {\bf 55},  5073
  (1997).

\bibitem{PoSt95}
S. Popp, O. Stiller, I. Aranson, and L. Kramer, Physica D {\bf 84},  398
  (1995).

\bibitem{BeFe67}
T. Benjamin and J. Feir, J. Fluid Mech. {\bf 27},  417  (1967).

\bibitem{CaEg97}
R. Cakmur, D. Egolf, B. Plapp, and E. Bodenschatz, Phys. Rev. Lett. {\bf 79},
  1853  (1997).

\bibitem{NaOt98}
K. Nam, E. Ott, M. Gabbay, and N. Guzdar, Physica D {\bf 118},  69  (1998).

\bibitem{SaRi99}
F. Sain and H. Riecke, Physica D, submitted  .

\end{thebibliography}

\end{document}